\documentclass[12pt,bibliography]{article}
\usepackage{cite}
\usepackage{lscape}
\usepackage{subfigure}
\usepackage{ytableau}
\usepackage[vcentermath]{youngtab}
\usepackage{cancel}
\usepackage{mathtools}
\usepackage{yfonts}
\usepackage{hyperref}
\usepackage{epsfig}
\usepackage{graphicx}
\usepackage{amsfonts} 
\usepackage{amsmath} 
\usepackage{color} 
\usepackage{pstool}
\usepackage[all]{xy}
\usepackage{xspace}

\newcommand{\Mathematica}{{\texttt{Mathematica}}\xspace}
\newcommand{\seed}{\emph{seed}\xspace}

\newcommand{\I}{\textrm{I}}
\newcommand{\II}{\textrm{II}}
\newcommand{\III}{\textrm{III}}
\newcommand{\IV}{\textrm{IV}}

\def\beq{\begin{equation}} 
\def\eeq{\end{equation}} 

\newcommand{\be}{\begin{equation}}
\newcommand{\ee}{\end{equation}}

\newcommand{\Acal}{{\mathcal A}}

\newcommand{\Dcal}{{\mathcal D}}

\newcommand{\Gcal}{{\mathcal G}}

\newcommand{\Mcal}{{\mathcal M}}
\newcommand{\Ncal}{{\mathcal N}}
\newcommand{\Ocal}{{\mathcal O}}
\newcommand{\Pcal}{{\mathcal P}}

\newcommand{\Scal}{{\mathcal S}}
\newcommand{\Tcal}{{\mathcal T}}

\newcommand{\Vcal}{{\mathcal V}}

\renewcommand{\tilde}{\widetilde}
\renewcommand{\hat}{\widehat}

\let\a=\alpha \let\b=\beta  \let\d=\delta 
    
\let\l=\lambda \let\m=\mu \let\n=\nu  
 \let\t=\tau    
  \let\D=\Delta

\begin{document}

\begin{titlepage}

{
\hfill
CERN-TH-2016-079
}
\vspace{1.5cm}
\begin{center}

{\Large \bf 
Radial expansion for spinning conformal blocks
}

\vspace{1cm}

{\bf Miguel S. Costa$^{1,2}$, Tobias Hansen$^1$, Jo\~ao Penedones$^{1,2,3}$, Emilio Trevisani$^1$}
\\
\vspace{1cm}
{\it  $^1$Centro de Fisica do Porto, Universidade do Porto, Porto, Portugal}
\\
\vspace{0.2cm} 
{\it $^2$Theory Division, Department of Physics, CERN,CH-1211 Gen\`eve 23, Switzerland}
\\
\vspace{0.2cm} 
{\it $^3$ Fields and Strings Laboratory, Institute of Physics, EPFL, CH-1015 Lausanne, Switzerland}

\end{center}

\vspace{1cm}

\begin{abstract}
This paper develops a method to compute any bosonic conformal block as a series expansion in the optimal radial coordinate introduced by Hogervorst and Rychkov. 
The method reduces to the known  result when the external operators are all the same scalar operator, but it allows to compute conformal blocks for external operators with spin. Moreover, we explain how to write closed form recursion relations for the coefficients of the expansions.
We study three examples of four point functions in detail: one vector and three scalars; two vectors and two scalars; two spin 2 tensors and two scalars.
Finally, for the case of two external vectors, we also provide a more efficient way to generate the series expansion using the analytic structure of the blocks as a function of the scaling dimension of the exchanged operator.

\end{abstract}

\bigskip

\end{titlepage}

\tableofcontents

\section{Introduction}
The  conformal bootstrap is opening a new window into  quantum field theory beyond perturbation theory
\cite{Ferrara:1973yt, Polyakov:1974gs, arXiv:0807.0004,
arXiv:0905.2211, arXiv:1009.2087, arXiv:1009.2725, arXiv:1009.5985, arXiv:1109.5176, arXiv:1203.6064, arXiv:1210.4258, arXiv:1304.1803, arXiv:1307.3111, arXiv:1307.6856, arXiv:1309.5089, arXiv:1403.4545, arXiv:1404.0489, arXiv:1406.4814, arXiv:1406.4858, arXiv:1412.4127, arXiv:1412.7541, arXiv:1502.02033, arXiv:1502.07217, arXiv:1503.02081, arXiv:1504.07997, arXiv:1507.05637, arXiv:1508.00012, arXiv:1510.03866,  arXiv:1511.04065, arXiv:1511.07552, arXiv:1511.08025, arXiv:1601.03476,
arXiv:1602.02810, arXiv:1603.03771, arXiv:1603.04436}.
In this paper we present a method to construct any bosonic conformal block (CB), an  essential ingredient to formulate the conformal bootstrap equations.

Constructing general CBs is still an open problem.
The exceptions are for three \cite{arXiv:1511.01497} and four \cite{arXiv:1601.05325} dimensions. \footnote{
In two dimensions, the CBs associated to the global conformal group are known in closed form \cite{arXiv:1205.1941}.}
 In these cases it was possible to construct the full set of \seed blocks which  are sufficient to determine all CBs by acting with differential operators \cite{arXiv:1109.6321,arXiv:1505.03750,arXiv:1508.00012}. 
However for general spacetime dimensions the problem is far from being solved. 
In fact, when the external operators have spin and the exchanged operator is not in the symmetric traceless representation of $SO(d)$ there is no systematic way to determine the blocks. \footnote{Results in this direction can be obtained using the shadow formalism. In particular,  \cite{arXiv:1411.7351,arXiv:1508.02676} considered  the CBs for two external scalars and two vectors. However this method requires  further systematization in order to address more general cases.}

The paper is divided into two parts.
The main goal is to explain how to generalize the series expansion in the radial coordinate introduced by \cite{Hogervorst:2013sma} for any bosonic conformal block in general dimension. 
This gives rise to a simple way to build CBs in a form which is suitable to the numerical bootstrap approach. Moreover it unravels their connection with the structure of the exchanged conformal representations. 
 In the text, we restrict our attention to external operators in the symmetric traceless representation of $SO(d)$ but this can be easily generalized using the methods of \cite{arXiv:1411.7351}. 
In section \ref{ScalarReview} we review the result of \cite{Hogervorst:2013sma} for the scalar CB (we actually consider the slightly more general case of scalar operators
of different dimension).  As a new result, we explain a strategy to obtain a closed form recurrence relation for the coefficients of the expansion. In section \ref{SpinningExpansion} we generalize this technology in order to deal with external operators with spin. We then exemplify the method in three cases: one external vector and three scalars, two external vectors and two scalars and two spin 2 operators and two scalars. 
 Moreover, we explain how to obtain recurrence relations for the coefficients of the expansions in general cases, and how to simplify these recurrence relation in order to make them more efficient (see also appendix \ref{diffeqingegbasis} on this point). 
 For clarity of exposition we give examples of simplified closed form recurrence relations only for the two cases of one external vector and three scalars and two external vectors and two scalars.
 For concreteness we also added to the arXiv submission some \Mathematica files which compute the blocks using the recurrence relations for the coefficients.

In section \ref{RecursionInDelta} we use an alternative method to find a radial expansion for CBs, by reconstructing their analytic structure as a  function of the conformal dimension of the exchanged operator. This method is based on an idea  of Zamolodchikov for two dimensional CFTs \cite{Zamolodchikov:1985ie}. It was recently developed in higher dimensions for the scalar conformal block \cite{arXiv:1307.6856,arXiv:1406.4858} and for two external fermions and two scalars in three dimensions  \cite{arXiv:1511.01497}. In \cite{ arXiv:1509.00428} there was a first attempt to generalize this idea to general external bosonic operators in any dimension. However there was still no explicit computation of CBs for the exchange of a mixed symmetry operator. 
In section \ref{RecursionInDelta}, and also in appendix \ref{appendix:recdeltaFULL}, we compute the CBs for two external vectors and two scalars for the exchange of both the symmetric and traceless representation, and the mixed symmetry representation 
$(l,1)=$\scalebox{0.23}{
\ytableausetup{centertableaux,boxsize=2.7 em}
$
\begin{ytableau}
\null & & &&   
  \none[\,\bullet \bullet \bullet] & \\
\\
\end{ytableau}
$} . Again we submitted a \Mathematica file which computes the blocks with this method.
On one hand this result can be used to match the radial expansion of section \ref{SpinningExpansion}. More importantly it is a concrete example that shows that the technology of \cite{ arXiv:1509.00428} can  actually be used to generate in principle any bosonic conformal block. 

Finally we conclude the paper with some comments on the possible strategies to obtain more general CBs. 
We also added appendix \ref{Spinning_CB} to explain how the idea of generating 
some (not all) CBs  for external operators with spin by acting with derivatives on the scalar \seed block \cite{arXiv:1109.6321} can be implemented in the radial coordinates 
of \cite{Hogervorst:2013sma}.

\section{Review of the scalar conformal block}
\label{ScalarReview}

\begin{figure}[h!]
\begin{centering}
\graphicspath{{Fig/}}
\def\svgwidth{5.9 cm} 
 \input{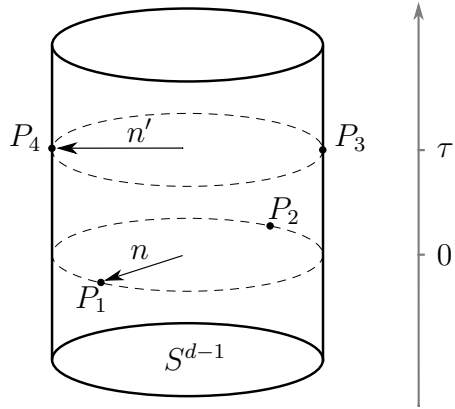}
\caption{\label{Cylinder_Configuration}
Cylinder configuration that leads to the radial coordinates of \cite{Hogervorst:2013sma}.
}
\end{centering}
\end{figure}

 Let us start by introducing a four point function of scalar operators written in the embedding space.
 We can expand it in conformal blocks as follows
\be
\langle \Ocal_1(P_1) \Ocal_2(P_2) \Ocal_3(P_3) \Ocal_4(P_4) \rangle = \sum_\Ocal c_{1 2 \Ocal } c_{3 4 \Ocal } G_{\D, l}(P_i) \ ,
\ee
where the sum is over primary operators $\Ocal$ with dimension $\D$ and spin $l$.
In the embedding space formalism the points are represented by null rays 
\be
P=\left(P^0,P^\mu,P^{d+1}\right)\in \mathbb{R}^{1,d+1}\ ,\qquad
P^2=-\left(P^0\right)^2+\left(P^{d+1}\right)^2+\delta_{\mu\nu}P^\mu P^\nu=0\ .
\ee
We will focus on the section $\delta_{\mu\nu}P^\mu P^\nu=1$. This is naturally parametrized by an Euclidean time $\tau \in \mathbb{R}$ and a unit vector $n^\mu \in S^{d-1} \subset \mathbb{R}^d$,
\be
P=(\cosh \tau,n^\mu, \sinh \tau)\ .
\ee
The induced metric on this section is just the metric on the cylinder  $\mathbb{R}\times S^{d-1}$, with line element
$
ds^2=d\tau^2+d\Omega^2_{S^{d-1}}
$.
The following choice of configuration   leads to the radial coordinates of \cite{Hogervorst:2013sma},
\begin{align}
\begin{split}
\label{PiOnTheCylinder}
&P_1=(1,n,0)\ ,\qquad \qquad \; \; \;
P_3=(\cosh \t,-n', \sinh \t)\ ,\\
&P_2=(1,-n,0)\ ,\qquad \qquad
P_4=(\cosh \t, n',\sinh \t)\ ,
\end{split}
\end{align}
 with $n$ and $n'$ unit vectors in $\mathbb{R}^d$. 
This choice of null vectors in the embedding space corresponds to the configuration on the cylinder $\mathbb{R}\times S^{d-1}$
shown in figure \ref{Cylinder_Configuration}.

Substituting the choice  (\ref{PiOnTheCylinder}) for $P_i$ in $G_{\D,l}$ we obtain the conformal block on the cylinder, 
\be
c_{1 2 \Ocal } c_{3 4 \Ocal } G_{\D,l}(P_i) \rightarrow \Gcal_{\D,l}(r, \eta) \ .
\ee
where we introduced the variables $r$ and $\eta$ of \cite{Hogervorst:2013sma}, 
\be
 r=e^{-\t} \ , \qquad \eta=n\cdot n' \ .
\ee
It is trivial to check that $r$ and $\eta$ are related to the usual cross ratios  as follows
\be
u=\frac{P_{12} P_{34}}{P_{13} P_{24}}= 
\frac{16 r^2}{\left(1+2 \eta  r+r^2\right)^2} \ , 
\qquad 
v= \frac{P_{14} P_{23}}{P_{13} P_{24}}=\frac{\left(1-2 \eta  r+r^2\right)^2}{\left(1+2 \eta  r+r^2\right)^2} \ ,
\ee
where $P_{ij}=-2P_i \cdot P_j$.

\subsection{A natural series expansion}
\begin{figure}[b!]
\begin{centering}
\graphicspath{{Fig/}}
\def\svgwidth{ 10 cm} 
 \input{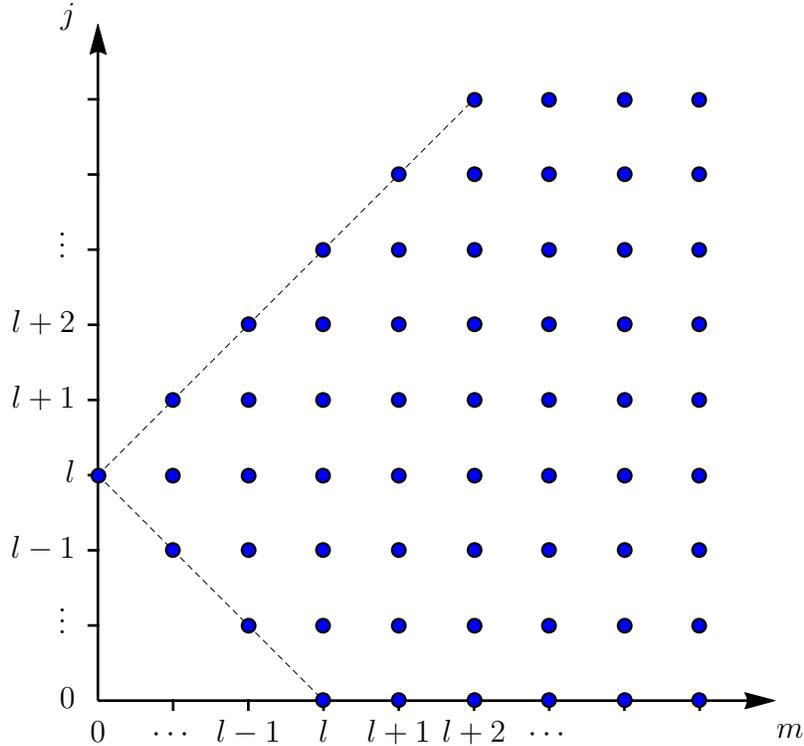}
\caption{\label{fig:Expansion_Plot} Support  of the coefficients $w(m,j)$ in the expansion of $\Gcal_{\D, l}$.
}
\end{centering}
\end{figure}

On the cylinder it is also possible to write
\be 
\label{GcalProjector}
  \Gcal_{\D, l}(r,\eta)
=
\langle
\mathcal{O}_4(n')
\mathcal{O}_3(-n')
|
r^{H_{cyl}} \mathcal{P}_l
|
\mathcal{O}_2(-n)
\mathcal{O}_1(n)
\rangle
\ ,
\ee
where $H_{cyl}$ is the Hamiltonian on the cylinder and $\mathcal{P}_l$ is the projector into the conformal family with primary $\mathcal{O}_{\Delta,l}$. 
It is natural to rewrite the projector as a sum over a complete set of states 
\ytableausetup{centertableaux,boxsize=1.2 em}
\begin{align}
\begin{split}
 \Gcal_{\D, l}(r,\eta)=  \sum_{m=0}^\infty r^{\Delta+m}\sum_{j={\rm max}(0,l-m)}^{l+m}
\sum_a 
\ &
\langle
\mathcal{O}_4(n')
\mathcal{O}_3(-n')
| m,{ \scriptsize
\begin{ytableau}
\mu_1&\mu_2&\, _{\cdots}&\mu_j \\
\end{ytableau},a
}\rangle
\label{eq:sumoverstates}
\\
& 
  \  \langle m,{ \scriptsize
\begin{ytableau}
\mu_1&\mu_2&\, _{\cdots}&\mu_j \\
\end{ytableau},a
} 
|
\mathcal{O}_2(-n)
\mathcal{O}_1(n)
\rangle \ ,
\end{split}
\end{align}
where the state $| m,{ \scriptsize
\begin{ytableau}
\mu_1&\mu_2&\, _{\cdots}&\mu_j \\
\end{ytableau} \;
},a \rangle$ represents descendants of the operator $\Ocal$ at the level $m$ and with spin $j$. The label $a$ distinguishes states at the same level and with the same spin.
In this case, the only states that contribute are symmetric traceless representations of $SO(d)$ because the inner products in (\ref{eq:sumoverstates}) vanish for more general representations.
 The inner product can be written as
\be
 \langle m,{ \scriptsize
\begin{ytableau}
\mu_1&\mu_2&\, _{\cdots}&\mu_j \\
\end{ytableau}
} ,a
|
\mathcal{O}_2(-n)
\mathcal{O}_1(n)
\rangle =  
u(m,j,a) \; n_{(\m_1} \cdots n_{\m_l)} \ ,
\ee
where the parenthesis select only the  traceless and symmetric part of the tensor.
We may choose to set $u(0,l)=c_{12\Ocal}$.
\footnote{At level $m=0$ there is no degeneracy, so we can drop the label $a$.}
From these definitions we automatically get
\be
\label{ScalarExpansionInR}
 \Gcal_{\D, l}(r,\eta)= \sum_{m=0}^\infty   r^{\Delta+m}
 \sum_{j={\rm max}(0,l-m)}^{l+m} w(m,j) \; \mathcal{C}_j(n \cdot n\rq{})\,,
\ee
where $w(m,j)\equiv \sum_a u(m,j,a) \tilde u(m,j,a)$, and $\tilde u(m,j,a)$ comes from the inner product involving  operators $\Ocal_3$ and $\Ocal_4$.
The function  $\mathcal{C}_j$ is a scalar spherical harmonic defined by
\be
\mathcal{C}_j(n\cdot n')=
 n_{ \a_1}\dots n_{\a_j}
\pi\big(
{ \scriptsize
\begin{ytableau}
\a_1&\a_2&\, _{\cdots}&\a_j \\
\end{ytableau}
}
,
{ \scriptsize
\begin{ytableau}
\b_1&\b_2&\, _{\cdots}&\b_j \\
\end{ytableau}
}
\big) 
 n'_{ \b_1}\dots n'_{\b_j} \ ,
\ee
where $\pi$ is the projector  into traceless and symmetric tensors with $j$ indices.
Moreover, the harmonic function $\mathcal{C}_l(\eta)$ can be explicitly written in terms of Gegenbauer polynomials,
\be
\mathcal{C}_l(\eta)\equiv \frac{l! }{(2 h-2)_l} \,C_l^{(h-1)}(\eta) \,,
\ee
where $h=d/2$.
The crucial point that makes the expansion (\ref{ScalarExpansionInR}) useful is that the coefficients $w(m,j)$ inherit some properties from the structure  the conformal representation exchanged in the conformal block.  For example, we automatically get that at level $m=0$ in the expansion there exists a single non-zero coefficient.
It corresponds to the exchange of the primary, namely
\be \label{InitialConditionScalar}
w(0,j)=  c_{1 2 \Ocal } c_{3 4 \Ocal } \d_{j,l} \ .
\ee
Moreover, all level $m$ descendants  must have spin $j$ in the interval $[{\rm max}(0,l-m),l+m]$, since they are created by the action of  $m$ derivatives, as each one of them can at most increase or decrease the spin of the operator by one. Therefore the coefficients have the following structure
\be
w(m,j)=0\,, \qquad |j-l|>m \,,
\ee
which is depicted in figure \ref{fig:Expansion_Plot}.
In addition, if $\Ocal_4=\Ocal_1$ and $\Ocal_3=\Ocal_2$ then unitarity implies that all coefficients are non-negative, $w(m,j)\ge 0$.

\subsection{A recurrence relation for the expansion coefficients}
We described the radial expansion for the conformal blocks, but we still need to fix the coefficients $w(m,j)$. In the following we explain how to use the conformal Casimir equation to obtain a recurrence relation for $w(m,j)$. To do so we first exploit the conformal symmetry to fix the conformal partial wave in terms of a function
depending only on the cross ratios, 
\be \label{CBScalarGtog}
G_{\D, l}(P_i)=  \frac{ 
\left(\frac{P_{24}}{P_{14}} \right)^{\frac{\D_1-\D_2}{2}}  \left(\frac{P_{14}}{P_{13}} \right)^{\frac{\D_3-\D_4}{2}}}{(P_{12})^{\frac{\D_1+\D_2}{2}}(P_{34})^{\frac{\D_3+\D_4}{2}}} \,g_{\D, l}(r,\eta) \ .
\ee
We then write the conformal Casimir equation in the embedding space, 
\be
(J_1+J_2)^2 G_{\D, l}(P_i)= c_{\D,l} G_{\D, l}(P_i)\ ,
\ee
with
\be
c_{\D,l}=\Delta  (\Delta -2 h)+l (2 h+l-2)\,,
\qquad 
J_k^{M N}=- i ( P_k^{A} \partial^{B}_{P_k} -P_k^{B} \partial^{A}_{P_k})  \,,
\ee
 in such a way to obtain a second order differential equation for $g_{\D, l}(r,\eta)$. To constrain the coefficients $w(m,j)$ we finally need to relate $\Gcal_{\D, l}(r,\eta)$ with $g_{\D, l}(r,\eta)$. This is trivial, we simply have to  write (\ref{CBScalarGtog}) using the definitions  (\ref{PiOnTheCylinder}) for the points $P_i$ on the cylinder,
\be 
\label{GcalTog}
\Gcal_{\D, l}(r,\eta)= c_{1 2 \Ocal } c_{3 4 \Ocal } \Pcal(r,\eta) g_{\D, l}(r,\eta) \,,
\ee
where
\be
\label{def:Pcal(P_i)}
 \frac{ 
\left(\frac{P_{24}}{P_{14}} \right)^{\frac{\D_1-\D_2}{2}}  \left(\frac{P_{14}}{P_{13}} \right)^{\frac{\D_3-\D_4}{2}}}{(P_{12})^{\frac{\D_1+\D_2}{2}}(P_{34})^{\frac{\D_3+\D_4}{2}}}
\quad \rightarrow \quad 
\Pcal(r,\eta) \equiv 
\frac{1}{2^{\sum_i\D_i}}
 \left(\frac{r^2+2 \eta  r+1}{r^2-2 \eta  r+1}\right)^{\frac{1}{2} \left(\Delta _{12}-\Delta _{34}\right)} \ .
\ee
Using these definitions we obtain a differential equation on $\Gcal_{\D, l}$,
\begin{align}
&\left(r^2-1\right)  \Big[f_1^2 f_2^2 c_{\D, l}
-4 r \left(\left(\Delta _{12}^2+\Delta _{34}^2\right) f_3+\Delta _{12} \Delta _{34} f_4 \eta  \left(r^2+1\right)\right)\Big] \Gcal_{\D, l}
\nonumber \\
&
+f_1 f_2 \eta  \left(r^2-1\right) \left(f_4+8 \eta ^2 h r^2-2 h \left(r^2+1\right)^2\right)
\partial_\eta \Gcal_{\D, l}
-f_1^2 f_2^2 \left(\eta ^2-1\right) \left(r^2-1\right) \partial_\eta^2 \Gcal_{\D, l}
 \label{ScalarCasimirEq}\\
&
-f_1 f_2 r \left[4 \eta ^2 r^2 \left(-2 h \left(r^2+1\right)+r^2+3\right)+\left(r^2+1\right) \left(2 h \left(r^2+1\right)^2+r^4-8 r^2-1\right)\right] \partial_r \Gcal_{\D, l}
\nonumber \\
&
-f_1^2 f_2^2 r^2 \left(r^2-1\right) 
\partial_r^2 \Gcal_{\D, l} = 0 \ ,
\nonumber
\end{align}
where
\begin{align}
\begin{split}
&f_1=(1 + r^2 - 2 r \eta) \ ,
 \qquad \qquad \qquad  \qquad  \ \  f_2=(1 + r^2 + 2 r \eta)\ ,\\
&f_3=r \left(\left(r^2+1\right)^2-2 \eta ^2 \left(r^4+1\right)\right) \ ,
 \qquad  f_4=r^2 \left(4 \eta ^2+r^2-6\right)+1 \ .
 \end{split}
\end{align}

We now explain how to convert any Casimir differential  equation into closed form recurrence relations for the coefficients of the expansion of the conformal blocks. 
One can convince oneself that any Casimir equation can be casted in the form\footnote{Notice that the Casimir equations may contain rational functions of $r$ and $\eta$. This is not a problem since one can always multiply the full expression by all the denominators.}
\be
\label{diffeqgeneric}
 p(r,\eta,\partial_r,\partial_\eta) \mathfrak F(r,\eta)=0\,,
\ee
where $p(x_1,x_2,x_3,x_4)$ is a polynomial in the variables $x_i$ and we choose the order of the variables such that the derivatives are on the right. As an example, \eqref{ScalarCasimirEq} is of the form \eqref{diffeqgeneric}. We want to show that if we expand $ \mathfrak F(r,\eta)$ as follows
\be
 \mathfrak F(r,\eta)=\sum_{m,j=0}^\infty  a_{m,j} \mathfrak f_{m,j}(r,\eta) \,, \qquad \mathfrak f_{m,j}(r,\eta)= r^{m} C_j(\eta)\,,
\ee
we are able to recast the differential equation (\ref{diffeqgeneric}) as a closed form algebraic relation on the coefficients $a_{m,j}$. 
To do so we need to know the action of $\eta$, $\partial_\eta$, $r$ and $\partial_r$ on the basis $ \mathfrak f_{m,j}(r,\eta)$, 
that is
\begin{align}
r \, \mathfrak f_{m,l}(r,\eta)&= \mathfrak f_{m+1,l}(r,\eta) \ , 
\qquad\qquad
\partial_r \, \mathfrak f_{m,l}(r,\eta)= m  \mathfrak f_{m-1,l}(r,\eta) \ ,
\nonumber\\
\eta  \, \mathfrak f_{m,l}(r,\eta)&=\frac{1}{2 (h+l-1)} \big( l \;\mathfrak f_{m,l-1}(r,\eta)+(2 h+l-2)  \mathfrak f_{m,l+1}(r,\eta) \big)\ ,
\label{action_r_eta}\\
\partial_\eta \, \mathfrak f_{m,l}(r,\eta)&=\frac{1}{\left(1-\eta ^2\right)} \frac{l (2 h+l-2) }{2 (h+l-1)}\big( \mathfrak f_{m,l-1}(r,\eta) -  \mathfrak f_{m,l+1}(r,\eta) \big) \ .
\nonumber
\end{align}
Using these formulas recursively one can  obtain the action of any polynomial $ p(r,\eta,\partial_r,\partial_\eta) $ on $\mathfrak f_{m,l}$.
The only subtlety comes from the derivatives in $\eta$, since they bring a factor $(1-\eta ^2)$ in the denominator. This is not a problem since we can multiply the full equation by $(1-\eta ^2)$ each time we use the last equation of (\ref{action_r_eta}).
This is enough to show that we can rewrite (\ref{diffeqgeneric}) as a closed form algebraic relation for the coefficients $a_{m,j}$. Nevertheless, in appendix \ref{diffeqingegbasis} we explain useful manipulations to simplify the algebraic relations.

\begin{figure}
\begin{centering}
\graphicspath{{Fig/}}
\def\svgwidth{ 10 cm} 
 \input{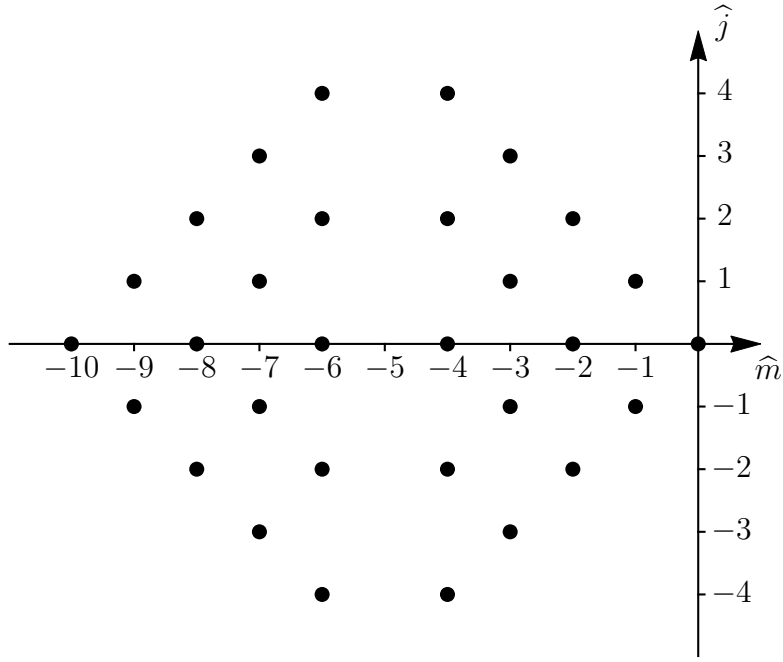}
\caption{\label{fig:RecCoefficientScalar} The set of points $\Scal$ involved in the recurrence relation for the coefficients $w(m,j)$ of the scalar conformal block. 
}
\end{centering}
\end{figure}

Applying this technology to the differential equation \eqref{ScalarCasimirEq} we obtain that a linear combination of a finite number of coefficients $w(m,j)$ with some shifts in $m$ and $j$ is zero, schematically
\be \label{recrelScalar}
\sum_{(\hat{m},\hat{j}) \in \Scal} c(\hat{m},\hat{j}) \; w(m+\hat{m},j+\hat{j})=0 \ .
\ee
 The set $\Scal$ contains $30$ points 
for example $(0,0),(-1,1),(-1,-1),\dots$, as pictured in figure \ref{fig:RecCoefficientScalar}, and $c(\hat{m},\hat{j})$ are known functions of the variables $\D_{12}$, $\D_{34}$, $\D$, $l$, $h$, and $m$ and $j$. We can therefore express the coefficient $w(m,j)$ with maximal $m$ (the point at the origin of the axes in figure \ref{fig:RecCoefficientScalar}) as a linear combination of $w(m\rq{},j\rq{})$ with lower $m\rq{}$. 

Moreover, if we set $m=0$ in   (\ref{recrelScalar})  we obtain a  constraint on the possible initial conditions 
\be
(j-l)  w(0,j)=0 \ ,
\ee
which is solved imposing $w(0,j)=0$ for $j \neq l$.
For $j = l$, instead, the coefficient $w(0,l)$ is left unrestricted. This is consistent with the initial condition (\ref{InitialConditionScalar}) obtained from the structure of the conformal families. 

Notice also that the restriction on the initial condition, together with the shape of figure \ref{fig:RecCoefficientScalar}, imply that the coefficients of the expansion organize as pictured in figure \ref{fig:Expansion_Plot}, which was obtained from the properties of conformal families.

Equation  (\ref{recrelScalar}) with the initial condition (\ref{InitialConditionScalar})  is enough to determine recursively the coefficients $w(m,j)$ for any $m$ and $j$. We will not present here  explicitly the relation (\ref{recrelScalar}) since the formula is very long. Instead, it is defined in a \Mathematica file.
 As an example we present the first coefficients of the expansion for generic parameters  
 \begin{align}
\begin{split}
w(1,l-1)&= \frac{2 \Delta _{34} l (-\Delta +2 h+l-2)+\Delta _{12} l \left(2 \Delta +\Delta _{34}-4 h-2 l+4\right)}{(h+l-1) (-\Delta +2 h+l-2)}
\ ,\\
w(1,l+1)&= \frac{(2 h+l-2) \big(2 \Delta _{34} (\Delta +l)-\Delta _{12} \left(\Delta _{34}+2 (\Delta +l)\right)\big)}{(h+l-1) (\Delta +l)} \ .
\end{split}
\end{align}
Notice that they go to zero when $\Delta _{34}=0=\Delta _{12}$, as expected from the results of \cite{Hogervorst:2013sma}.

Formula (\ref{recrelScalar}) is a new result, since in \cite{Hogervorst:2013sma} it was not obtained a closed form recurrence relation for the coefficients of the expansion. The main reason for this is that the authors of \cite{Hogervorst:2013sma} kept some rational functions of $r$ and $\eta$ (which have infinite expansion in small $r$) in the Casimir differential equation. Thus the  associated recurrence relation would have had a set  $\Scal$ containing an infinite number of points. On the other hand,  multiplying the Casimir equation by all the denominators it becomes possible to find a recurrence relation with a finite set  $\Scal$. Notice that this example explicitly shows that there exist many possible equivalent recurrence relations, which however are not equivalently efficient. In fact we expand on this point in appendix \ref{diffeqingegbasis}, where we explain some strategies to further reduce the number of points in $\Scal$ with respect to the algorithm proposed in the main text.
\section{Spinning conformal blocks}
\label{SpinningExpansion}


In this section we present the general framework to obtain the radial expansion of conformal blocks for external bosonic operators with spin. Initially we exemplify the formalism applying it to the simpler case of one external vector operator and three scalars, for which the allowed exchanged operators are still symmetric and traceless. 
After that, in separate subsections, we apply the method  to the cases of two external vectors and two scalars, and of two spin 2 operators and two scalars. In these cases there are new mixed tensor operators that can be exchanged.

\subsection{General formalism and  example of one  vector and three scalars}
\label{subsec:1extvec}
We first introduce the conformal block expansion  in the embedding formalism of a four point function of operators  $\Ocal_i$ with spin $l_i$ and conformal dimension $\D_i$
\be \label{CBExpansionGeneral}
\langle \Ocal_1(P_1,Z_1)\Ocal_2(P_2,Z_2) \Ocal_3(P_3,Z_3)\Ocal_4(P_4,Z_4)\rangle = \sum_\Ocal \sum_{p,q} c^{(p)}_{12 \Ocal} c^{(q)}_{34 \Ocal} G^{(p,q)}_{\l}(P_i,Z_i)  \ ,
\ee
where the first sum runs over  all  the  primary operators $\Ocal$ in representations labeled by $\l=(\D,l_1,l_2,\dots,l_{[\frac{d}{2}]})$, where $[x]$ rounds $x$ down to the closest integer. We also sum over $p$ and $q$, which label the possible conformal invariant tensor structures of the three point functions
$\langle {\cal O}_1 {\cal O}_2 {\cal O}\rangle$ and $\langle {\cal O}_3 {\cal O}_4 {\cal O}\rangle$, respectively. 
Let us recall that the auxiliary variables $Z_i$ are a convenient way to avoid dealing with tensor indices in embedding space, that is 
$\Ocal(P,Z)=Z^{A_1}\dots Z^{A_l} \Ocal_{A_1 \dots A_l}(P)$.
As an example, for one external vector we have
\be
\label{vectorCB}
\langle \Ocal_1(P_1,Z_1)\Ocal_2(P_2) \Ocal_3(P_3)\Ocal_4(P_4)\rangle = \sum_\Ocal \sum_{p=1}^2 c^{(p)}_{12 \Ocal} c_{34 \Ocal} G^{(p)}_{\D, l}(P_i,Z_1)  \ .
\ee
The two possible values $p=1,2$ correspond to the two possible tensor structures of the three-point function $\langle \Ocal_1(P_1,Z_1)\Ocal_2(P_2)\Ocal(P,Z)\rangle$ between external operators $\Ocal_1$, $\Ocal_2$ and the exchanged operator $\Ocal$, which in this example can only be symmetric traceless  of generic spin $l$.
As in the scalar case, we are interested in writing the conformal blocks in the cylinder configuration. To do so we can use formulae (\ref{PiOnTheCylinder}) and the following definition for  the embedding space polarization vectors
\begin{align}
\label{ZiOnTheCylinder}
\begin{split}
&Z_1=(z_1\cdot n,z_1,z_1\cdot n)\ ,\qquad \qquad \quad \;\ 
Z_3=(-z_3\cdot n' e^\t, z_3,- z_3\cdot n' e^\t)\ ,\\
&Z_2=(-z_2\cdot n,z_2,-z_2\cdot n)\ ,\qquad \qquad
Z_4=(z_4\cdot n' e^\t, z_4, z_4\cdot n' e^\t)\ ,
\end{split}
\end{align}
where $z_i$ are polarization vectors in $\mathbb{R}^d$ which obey $z_i^2=0$. Notice that $Z_i^2=Z_i\cdot P_i=0$, as it should \cite{arXiv:1107.3554}. With these definitions we get 
\be
\sum_{p,q} c^{(p)}_{12 \Ocal} c^{(q)}_{34 \Ocal}G_{\l}^{(p,q)}(P_i,Z_i) \rightarrow \Gcal_{\l}(r, \eta , n\cdot z_i,n'\cdot z_i,z_i\cdot z_j) \ .
\ee
Therefore $\Gcal_{\l}^{(p,q)}$ can be expanded in polynomials $p_{s}$ of the variables $n\cdot z_i,\ n'\cdot z_i,\ z_i\cdot z_j$ with weight $l_i$ in each variable $z_i$,
\be \label{GcalToFGeneric}
\Gcal_{\l}(r, \eta , n\cdot z_i,n'\cdot z_i,z_i\cdot z_j)=\sum_{s} p_{s}(n\cdot z_i,n'\cdot z_i,z_i\cdot z_j) F_s(r,\eta) \  ,
\ee
where the coefficient multiplying each polynomial is a function $F_s$ of $r$, and $\eta$
and $s$ labels the independent tensor structures of the four-point function. For one external vector we have
\be
\label{GacalToF1Vector}
\Gcal_{\D, l}=   (n\cdot z_1) F_1(r,\eta)+(n\rq{} \cdot z_1) F_2(r,\eta)  	\ .
\ee

On the other hand, just as for the scalar case (\ref{GcalProjector}),  the conformal block on the cylinder can be written as
\be \label{Gcalpq}
\Gcal_{\l}(r,\eta, n\cdot z_i,n'\cdot z_i,z_i\cdot z_j)
=
\langle
\mathcal{O}_4(n',z_4)
\mathcal{O}_3(-n',z_3)
|
r^{H_{cyl}} \mathcal{P}_\lambda
|
\mathcal{O}_2(-n,z_2)
\mathcal{O}_1(n,z_1)
\rangle
 \ ,
\ee
where  $\mathcal{P}_\lambda$ is the projector into the conformal family with highest weight $\l$.
It is natural to rewrite the projector as a sum over a complete basis of states 
\begin{align} \label{GcalSumStates}
&\Gcal_{\l} =  \sum_{m=0}^\infty r^{\Delta+m}
\sum_{\bf  Y}
\sum_a \langle
\mathcal{O}_4(n',z_4)
\mathcal{O}_3(-n',z_3)
| m,{\bf  Y},a\rangle \langle m,{\bf  Y},a 
|
\mathcal{O}_2(-n,z_2)
\mathcal{O}_1(n,z_1)
\rangle
\,,
\end{align}
where we sum over all states at level $m$ of the conformal family, organized in irreducible representations (irreps) ${\bf  Y}$ of $SO(d)$.  
For example, for one external vector, the representation ${\bf  Y}$ can only be symmetric and traceless, therefore
\ytableausetup{centertableaux,boxsize=1.2 em}
\be
\langle m,{ \scriptsize
\begin{ytableau}
\mu_1&\mu_2&\, _{\cdots}&\mu_j \\
\end{ytableau}
},a
|
\mathcal{O}_2(-n)
\mathcal{O}_1(n,z_1)
\rangle= 
\big[ 
 u_1(m,j,a)\,z_1\cdot n +
 u_2(m,j,a)\, z_1\cdot \nabla_n 
\big]n_{(\mu_1}\dots n_{\mu_j)}\,,
\label{3pt:j10} 
\ee
where
\be
(\nabla_n)_\mu=\frac{\partial}{\partial n^\mu}-n_\mu n\cdot \frac{\partial}{\partial n}
\ee
 computes the covariant derivative on the sphere $S^{d-1}$ parametrized by $n$ and the coefficients $u_{k}$ are not fixed by rotational invariance.  
We can choose a basis for the OPE coefficients $c_{12\Ocal}^{(p)}$ such that 
\ytableausetup{centertableaux,boxsize=1.2 em}
\begin{align} \label{1vecinitialcond}
& \langle 0,{ \scriptsize
\begin{ytableau}
\mu_1&\mu_2&\, _{\cdots}&\mu_j \\
\end{ytableau}
}
|
\mathcal{O}_2(-n)
\mathcal{O}_1(n,z_1)
\rangle= 
\left[ 
 c^{(1)}_{12 \Ocal}\,z_1\cdot n +
c^{(2)}_{12 \Ocal}\, z_1\cdot \nabla_n 
\right]n_{(\mu_1}\dots n_{\mu_j)}
\,.
\end{align}
In this formalism we can easily deal with more complicated  irreps  ${\bf  Y}$ of $SO(d)$, which will appear in the next sections. For now, we continue the study of the case with one external vector, for which (\ref{GcalSumStates}) becomes
\be \label{GcalExpansion1Vector}
  \Gcal_{\D, l}= \sum_{m=0}^\infty r^{\D+m}\sum_{j } \sum_a \tilde u(m,j,a) \big[ 
u_1(m,j,a)\,z_1\cdot n +
u_2(m,j,a)\, z_1\cdot \nabla_n 
\big]
 \mathcal{C}_j(n \cdot n\rq{}) \ ,
\ee
where $\tilde u(m,j,a)$ comes from the inner product of the scalars ${\cal O}_3$ and ${\cal O}_4$ with the exchanged states.
It is convenient to introduce the following  two functions of $r$ and $\eta$,
\be
W_{s}(r,\eta)\equiv \sum_{m=0}^\infty 
r^{\D+m} 
\sum_{j={\rm max}(0,l-m)}^{l+m} w_{s}(m,j)  \mathcal{C}_j(\eta) \ , \qquad
\qquad s=1,2\ ,
\ee
where $w_{s } (m,j)\equiv \sum_a \tilde u(m,j,a) u_{s }(m,j,a) $. 
The choice \eqref{1vecinitialcond} corresponds to the following initial condition 
\be
\label{InitialCondition1Vec}
w_{s} (0,j)=  c_{12\Ocal}^{(s)} c_{34\Ocal} \d_{j,l}  \ .
\ee

The functions $W_s$ are the basic building blocks for the radial expansion. However they are related in a non trivial way to the full partial wave $G_\l$ in (\ref{CBExpansionGeneral}). In the following we explain how to relate these functions in a generic framework while always exemplifying it for the case of one external vector operator.

As a first step we relate the functions $W_s$ to the functions $F_s$ defined in \eqref{GcalToFGeneric}.
For one external vector this can be done expanding
\eqref{GcalExpansion1Vector} and collecting terms according to  (\ref{GacalToF1Vector}),
\be
F_1(r,\eta)=W_1(r,\eta)- \eta \partial_\eta W_2(r,\eta) \ ,\qquad \qquad
F_2(r,\eta)=\partial_\eta W_2(r,\eta) \ .
\label{FExpansion}
\ee

The second step is to relate the conformal partial waves  $G_\l$ to the functions $F_s$. To do so we exploit conformal  symmetry to write the full partial wave in terms of functions of the cross ratios. This can be easily done for any conformal block in the embedding formalism,
\be 
G_{\l}^{(p,q)}(P_i,Z_i)=\frac{ 
\left(\frac{P_{24}}{P_{14}} \right)^{\frac{\D_1-\D_2}{2}}  \left(\frac{P_{14}}{P_{13}} \right)^{\frac{\D_3-\D_4}{2}}}{(P_{12})^{\frac{\D_1+\D_2}{2}}(P_{34})^{\frac{\D_3+\D_4}{2}}} \sum_{s} g^{(p,q)}_{\l,s}(r, \eta) Q^{(s)}(P_i,Z_i)\ , \label{CB:structuresEmbedding}
\ee
where the set of $Q^{(s)}(P_i,Z_i)$ encode the conformal invariant tensor structures and can be generated using \cite{arXiv:1107.3554,arXiv:1411.7351}
\begin{align} 
\begin{split}
V_{i,jk} & \equiv -\frac{(P_i\cdot P_j) (P_k\cdot Z_i)-(P_i\cdot P_k) (P_j\cdot Z_i)}{ \sqrt{-2 \left(P_i\cdot P_j\right) (P_i\cdot P_k) (P_j\cdot P_k)}} \ , \\
H_{ij} & \equiv \frac{(P_i\cdot P_j) (Z_i\cdot Z_j)-(P_i\cdot Z_j) (P_j\cdot Z_i)}{(P_i\cdot P_j)} \ .
\end{split}
\end{align}
Finally we map (\ref{CB:structuresEmbedding}) to the cylinder using the definitions (\ref{PiOnTheCylinder}) and (\ref{ZiOnTheCylinder}), and we collect terms according to \eqref{GcalToFGeneric}. In this way we obtain a linear relation between the functions $g_{\l,s}$ and $F_s$.

In the one vector case we only have two possible tensor structures in (\ref{CB:structuresEmbedding}),
\be
Q^{(1)}(P_i,Z_i)= V_{1,23} \ , \qquad Q^{(2)}(P_i,Z_i)= V_{1,43} \ ,
\ee
which can be evaluated on the cylinder using  (\ref{PiOnTheCylinder}) and (\ref{ZiOnTheCylinder}),
\begin{align}
\begin{split}
\label{VsOnTheCylinder}
V_{1,23}& \to
\frac{\left(-r^2+2 \eta  r+1\right) n\cdot z_1-2 r z_1\cdot n'}{(r^2-2 \eta  r+1)^{1/2}(r^2+2 \eta  r+1)^{1/2}} \ , \\
V_{1,43}& \to \frac{2 \eta  r^2 n\cdot z_1-\left(r^2+1\right) z_1\cdot n'}{(r^2-2 \eta  r+1)^{1/2}(r^2+2 \eta  r+1)^{1/2}} \ . 
\end{split}
\end{align}
We then replace \eqref{VsOnTheCylinder} and \eqref{def:Pcal(P_i)} in (\ref{CB:structuresEmbedding}). Collecting terms according to \eqref{GacalToF1Vector} we finally get
\begin{align}
\label{gtoF}
\begin{split}
\sum_{p=1}^2 c^{(p)}_{12 \Ocal} c_{34 \Ocal}g^{(p)}_1(r,\eta)&=\Acal(r,\eta ) \left[2 \eta  r^2 F_2(r,\eta )+\left(r^2+1\right) F_1(r,\eta )\right] ,
\\
\sum_{p=1}^2 c^{(p)}_{12 \Ocal} c_{34 \Ocal}g^{(p)}_2(r,\eta)&=\Acal(r,\eta ) \left[\left(r^2-2 \eta  r-1\right) F_2(r,\eta )-2 r F_1(r,\eta )\right] , 
\end{split}
\end{align}
where we defined the function 
\be
\Acal(r,\eta)=-\frac{2^{\sum_i \D_i}}{1 - r^2}
\left(\frac{1-2 \eta  r+r^2}{1+2 \eta  r+r^2}\right)^{\frac{1}{2} \left(\Delta _{12}-\Delta _{34}+1\right)}\ .
\ee

\begin{figure}[t!]
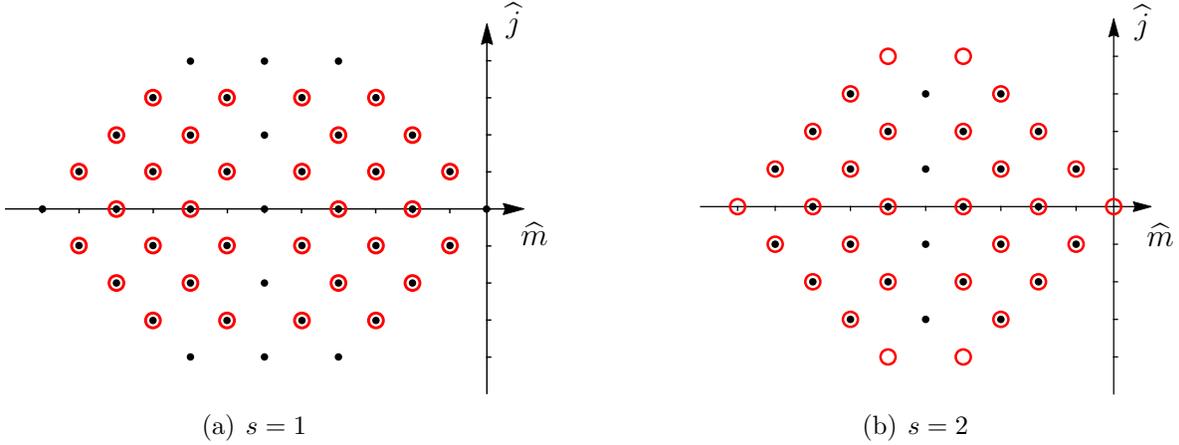

\centering
\subfigure[\label{fig:RecCoefficient1Vector1} $s= 1$]{\graphicspath{{Fig/}}
\def\svgwidth{6.9 cm} 
 \input{RecCoefficient1Vector1.pdf_tex}}
~~~~~~~~~~~~~
\subfigure[\label{fig:RecCoefficient1Vector2} $s= 2$]
{\graphicspath{{Fig/}}
\def\svgwidth{6 cm} 
 \input{RecCoefficient1Vector2.pdf_tex}}
\caption{ The sets $\Scal^s_{s\rq{}}$ appearing in the linear combination (\ref{recrel1Vector}) for $s=1,2$.
Red circles correspond to $s'=2$ and black dots correspond to $s'=1$.
}
\end{figure}

With the above relations we can write $G_{\l}$ in terms of the functions $W_s$. 
The last step is to find a recurrence relation for the coefficients $w_s(m,j)$ of the functions $W_s$. To do so, we first write the conformal Casimir equation in the embedding space,
\begin{align}
&(J_1+J_2)^2 G_{\l} (P_i,Z_i)=c_\l G_{\l} (P_i,Z_i) \ , \qquad c_\l=\D(\D-2h)+\sum_{i=1}^{[h]} l_i (l_i+2 h -2 i) \ ,
\label{eq:casimir}
\end{align}
where $J_i$ are the generators of conformal transformations 
\be
J_k^{M N}=-i \left( P_k^{M} \partial^{N}_{P_k} -P_k^{N} \partial^{M}_{P_k} +  Z_k^{M} \partial^{N}_{Z_k} -Z_k^{N} \partial^{M}_{Z_k} \right) \ .
\ee
We then write $G_{\l} $ in terms of $W_s$, which leads to coupled differential equations for the functions $W_s$.
Repeating the same procedure described in the scalar case, we can  trade (see also appendix \ref{diffeqingegbasis}) the differential equations for  coupled recurrence relations on the coefficients $w_s(m,j)$. When we do this for the case of one external vector we obtain two coupled recurrence relations of the kind (the precise formulae are defined in a \Mathematica file)
 \be \label{recrel1Vector}
\sum_{s'=1}^2 \sum_{(\hat{m},\hat{j}) \in \Scal^s_{s'}} c^s_{s'}(\hat{m},\hat{j}) \; w_{s'}(m+\hat{m},j+\hat{j})=0
 \ , \qquad s=1,2\ ,
\ee
where the sets of point $\Scal^s_{s'}$ are represented in figures  \ref{fig:RecCoefficient1Vector1} and \ref{fig:RecCoefficient1Vector2}.
The sets $\Scal^s_1$ are pictured by black dots while the set $\Scal^s_2$ by red circles. In particular the sets $\Scal^1_{s'}$ of the first recurrence relation are shown in figure \ref{fig:RecCoefficient1Vector1} while the sets $\Scal^2_{s'}$ of the second one are in figure \ref{fig:RecCoefficient1Vector2}.
We can then obtain the coefficients of the expansion iterating (\ref{recrel1Vector}) with the initial condition (\ref{InitialCondition1Vec}).
Equation (\ref{recrel1Vector}) for $m=0$  reduces to
\be
(j - l) w_1(0,j)=0\ ,  \qquad  j(j - l) w_2(0,j)=0 \ , \label{IC1spin1}
\ee
which constraints  the possible initial conditions. In fact,  \eqref{IC1spin1} implies $w_1(0,j)=w_2(0,j)=0$,  for any $j \neq l$.
On the other hand, $w_{1}(0,l)$ and $w_{2}(0,l)$ are not constrained by equation (\ref{recrel1Vector}).
 This is consistent with the initial condition (\ref{InitialCondition1Vec}) that we derived from the structure of the conformal families. The two independent solutions of equation (\ref{recrel1Vector}) are the two conformal blocks that we labeled with $p=1,2$. 
The last remark is that (\ref{recrel1Vector}) does not constrain $w_2(m,0)$. This actually make sense since in \eqref{FExpansion} the function $W_2$ only appears derived once in $\eta$ and therefore the $j=0$ term is annihilated. In fact, for practical purposes one could redefine $W_2$ avoiding to sum over the terms with $j=0$.

As an example we present all the coefficients at level $m=1$ of the expansion of the functions $W_s$, which build the conformal block $G^{(p)}_{\D, l}$. At level zero we set the initial condition $w_s(0,l)= \d_{s, p}$. Using \eqref{recrel1Vector} we therefore obtain
 \begin{align} 
\begin{split}
w_1(1,l+1)&=-\frac{\Delta _{34} (2 h+l-2) \big(\Delta _{12} \d_{p,1}+l \d_{p,2}\big) }{(h+l-1) (\Delta +l)}  \,, 
\\
w_2(1,l+1)&=-\frac{\Delta _{34} (2 h+l-2) \big(\d_{p,1}+\Delta _{12} l\d_{p,2}\big) }{(l+1) (h+l-1) (\Delta +l)} \,, 
\\
w_1(1,l-1)&=\frac{\Delta _{34} l \big(\Delta _{12} \d_{p,1} -(2 h+l-2) \d_{p,2}\big)}{(h+l-1) (-\Delta +2 h+l-2)} \,, \\
w_2(1,l-1)&=-\frac{\Delta _{34} l \big(\d_{p,1}-\Delta _{12} (2 h+l-2) \d_{p,2}\big)}{(h+l-1) (2 h+l-3) (-\Delta +2 h+l-2)}  \,.
\end{split}
\end{align}
\subsection{Two external vectors and two scalars}
In this section we describe how to obtain a radial expansion for conformal blocks of  two external vectors and two scalars. 
In particular, we consider the following conformal block decomposition 
\begin{align}
\begin{split}
\label{eq:4ptCBdecomposition}
\langle \Ocal_1(P_1,Z_1)\Ocal_2(P_2) \Ocal_3(P_3,Z_3)\Ocal_4(P_4)\rangle =
&  \sum_{\Ocal \in 
\ytableausetup{centertableaux,boxsize=0.3 em}
\begin{ytableau}
\null &&  & \\
\end{ytableau}
}
\; \sum_{p,q=1}^2 c^{(p)}_{12 \Ocal} c^{(q)}_{34 \Ocal} G^{(p,q)}_{\D, l}(P_i,Z_i)   \\
&+ \sum_{\Ocal \in 
\ytableausetup{centertableaux,boxsize=0.3 em}
\begin{ytableau}
\null &&  & \\
\\
\end{ytableau}
}
c_{12 \Ocal} c_{34 \Ocal} G_{\D, l , 1}(P_i,Z_i) \ .
\end{split}
\end{align}
There exist four conformal partial waves $G^{(p,q)}_{\D, l}$ (for $p,q=1,2$) when the exchanged multiplet has an highest weight $\Ocal$ in the  symmetric and traceless representation. Additionally, there is a single partial wave  $G_{\D ,l ,1}$ when $\Ocal$ is in the representation $(\D,l,1)$. We shall see that the radial expansion allows us to find all the cases at once.

First, we consider the contribution from each conformal family on the cylinder
\be
\sum_{p,q=1}^2 c^{(p)}_{12 \Ocal} c^{(q)}_{34 \Ocal} G^{(p,q)}_{\D, l}(P_i,Z_i) \to \Gcal_{\D, l}\ ,\qquad
c_{12 \Ocal} c_{34 \Ocal} G_{\D ,l , 1}(P_i,Z_i)
\to \Gcal_{\D, l , 1}\ .
\ee
On the cylinder, both these functions can be expanded 
in the following tensor structures 
\begin{align}
\begin{split}
\Gcal_{\l} = \
&
z_1\cdot n \, z_3\cdot n'\,F_1(r,\eta)+
z_1\cdot n \,z_3\cdot n\,F_2(r,\eta)+
\label{definitionofFs}
\\&
+z_1\cdot n' \,z_3\cdot n' \,F_3(r,\eta)+
z_1\cdot n' \,z_3\cdot n\,F_4(r,\eta)+
z_1\cdot z_3\,F_5(r,\eta)\,.
\end{split}
\end{align}
The functions $F_s$ depend on the representation $\l$ but we omit this label to avoid cluttering the equations.

To implement the expansion (\ref{GcalSumStates}) for two external vectors we need to consider the contribution from two types of states at each level $m$ (the descendants
of the two $SO(d)$ irreps that are exchanged in this case).
The first is the same that we obtained in the one external vector case \eqref{3pt:j10}. The second one  is new, with a coupling between one of the vectors and one of the scalars written as
\ytableausetup{centertableaux,boxsize=1.2 em}
\begin{align}
&\Big\langle m,{ \scriptsize
\begin{ytableau}
\mu_1&\mu_2&\, _{\cdots}&\mu_j \\
\nu
\end{ytableau}
},a
|
\mathcal{O}_2(-n)
\mathcal{O}_1(n,z_1)
\Big\rangle =N_\mathcal{V} \;
u_0(m,j,a)\,
\pi\Big(
{ \scriptsize
\begin{ytableau}
\mu_1&\mu_2&\, _{\cdots}&\mu_j \\
\nu
\end{ytableau}
}
,
{ \scriptsize
\begin{ytableau}
\a_1&\a_2&\, _{\cdots}&\a_j \\
\b
\end{ytableau}
}
\Big)
 (z_1)_\b 
 n_{ \a_1}\dots n_{\a_j} \ ,
\end{align}
where the normalization constant $N_\mathcal{V}$ is introduced for later convenience.
Notice that the value at $m=0$ of $u_{s}(0,j)$ has to be different depending on what is the representation $\l$ of the exchanged primary operator. 
In fact, if $\l$ is the symmetric and traceless representation, then 
\be \label{IC2vectorl}
u_0 (0,l)=0 \ , \qquad 
u_q (0,l)=c_{12\Ocal}^{(q)}   \ , \qquad 
q=1,2\ ,
\ee
and when $\l=(\D,l,1)$, then 
\be \label{IC2vectorl1}
u_0 (0,l)= c_{12\Ocal}\ ,  \qquad 
u_q(0,l)=0 \ , \qquad  
q=1,2 \ .
\ee
We are lead to the following expansion for the conformal block on the cylinder
\begin{align}
&\Gcal_{\l}= 
\sum_{m=0}^\infty r^{\Delta+m} 
\sum_j \sum_a
\Big[ N_\mathcal{V}^2 \; u_0(m,j,a) \tilde u_0(m,j,a) (z_1)_\mu (z_3)_\nu  \mathcal{V}_j^{\mu\nu}(n,n')+
\label{SphericalHarmonicExpansion}
\\
&+
\big( 
u_1(m,j,a)\,z_1\cdot n +
u_2(m,j,a)\, z_1\cdot \nabla_n 
\big)\big( 
\tilde u_1(m,j,a)\,z_3\cdot n' +
\tilde u_2(m,j,a)\, z_3\cdot \nabla_{n'} 
\big)  \mathcal{C}_j(n\cdot n') \Big] , 
\nonumber
\end{align}
where 
\begin{align}
 \mathcal{V}_j^{\mu\nu}(n, n')&=
 n_{ \a_1}\dots n_{\a_j}
\pi\Big(
{ \scriptsize
\begin{ytableau}
\a_1&\a_2&\, _{\cdots}&\a_j \\
\mu
\end{ytableau}
}
,
{ \scriptsize
\begin{ytableau}
\b_1&\b_2&\, _{\cdots}&\b_j \\
\nu
\end{ytableau}
}
\Big) 
 n'_{ \b_1}\dots n'_{\b_j} 
\end{align}
is a (double) vector spherical harmonic \footnote{See appendix F of the companion paper \cite{projectors} on the relation between projectors and spherical harmonics.} with index $\mu$ at the point $n$ and index $\nu$ at the point $n'$. The double vector harmonic can be expressed in terms of Gegenbauer functions as follows \cite{projectors}
\begin{align}
\label{Vmunu}
\begin{split}
 N_\mathcal{V}^2 \; \mathcal{V}_l^{\mu\nu}(n,n') (z_1)_\mu (z_3)_\nu =\ &\mathcal{C}_l''(\eta ) \left(-\eta  n\cdot z_1 z_3\cdot n'+z_1\cdot n' z_3\cdot n'+n\cdot z_1 n\cdot z_3-z_1\cdot z_3\right) \\
&-\left(n\cdot z_3 z_1\cdot n'-\eta  z_1\cdot z_3\right) \big((2 h-2) \mathcal{C}_l'(\eta )+\eta  \mathcal{C}_l''(\eta )\big) \,,
\end{split}
\end{align}
where the normalization constant $N_\mathcal{V}^2$ is fixed by requiring that $\pi$ is a projector
\be
\label{NormalizationNV}
N_\mathcal{V}^2 \equiv \frac{2^l l (h)_{l-1}}{(2 h-1)_{l-2}} \ .
\ee
It is then convenient to introduce five functions which have a nice expansion in radial coordinates and which encode the information to generate the conformal blocks
\be \label{eq:seriesWpq}
W_{s}(r,\eta)=\sum_{j,m} w _{s}(j,m) r^{\D+m} \mathcal{C}_j(\eta) \ , \qquad  s=1,2,3,4,5\,,
\ee 
with 
\be
\begin{array}{l}
w_{1 }(m,j) =
\sum_a u_{1}(m,j,a) \tilde u_{1}(m,j,a) \ , \\
w_{2 }(m,j) =
\sum_a u_{1}(m,j,a) \tilde u_{2}(m,j,a) \ , \\
w_{3 }(m,j) =
\sum_a u_{2}(m,j,a) \tilde u_{1}(m,j,a) \ , \\
w_{4 }(m,j) =
\sum_a u_{2}(m,j,a) \tilde u_{2}(m,j,a) \ , \\
w_{5 }(m,j) =
\sum_a u_{0}(m,j,a) \tilde u_{0}(m,j,a) \ .
\end{array}
\ee

Next we want to write the conformal partial waves in terms of the functions $W_{s}$. First we obtain the relation between $F_s$ and $W_{s}$ according to the definition \eqref{definitionofFs},
\begin{align}
 F_1(r,\eta )&=W_{1}(r,\eta )+\eta \partial _{\eta } \big[\eta 
   \partial _{\eta } W_{4}(r,\eta ) - W_{2}(r,\eta )-W_{3}(r,\eta )-\partial
   _{\eta } W_{5}(r,\eta )\big] \,,
   \nonumber\\
 F_2(r,\eta )&=\partial _{\eta }\big[- \eta  \partial _{\eta
   } W_{4}(r,\eta )+W_{2}(r,\eta )+\partial _{\eta }W_{5}(r,\eta ) \big] \,,
   \nonumber\\
 F_3(r,\eta )&=\partial _{\eta }\big[ -\eta  \partial _{\eta
   } W_{4}(r,\eta )+W_{3}(r,\eta )+\partial _{\eta } W_{5}(r,\eta ) \big] \,,
   \label{FtoW2vec}
\\
 F_4(r,\eta )&=-\big[2 (-1+h) +\eta  \partial _{\eta }\big] \partial _{\eta }W_{5}(r,\eta   )
  +\partial _{\eta }{}^2W_{4}(r,\eta ) \,,
  \nonumber\\
 F_5(r,\eta )&=\big[2 (-1+h) \eta  
+\left(-1+\eta
   ^2\right) \partial _{\eta } \big] \partial _{\eta }W_{5}(r,\eta
   )
+\partial _{\eta }W_{4}(r,\eta )\,.
  \nonumber
\end{align} 
Secondly we write the conformal partial wave  in the embedding space in terms of functions of the cross ratios
\be 
\label{GTog_Generic}
\sum_{p,q} c^{(p)}_{12 \Ocal} c^{(q)}_{34 \Ocal}G_{\l}^{(p,q)}(\{P_i,Z_i\})=\frac{ 
\left(\frac{P_{24}}{P_{14}} \right)^{\frac{\D_1-\D_2}{2}}  \left(\frac{P_{14}}{P_{13}} \right)^{\frac{\D_3-\D_4}{2}}}{(P_{12})^{\frac{\D_1+\D_2}{2}}(P_{34})^{\frac{\D_3+\D_4}{2}}} \sum_{s=1}^{5} g_{\l,s}(r, \eta) Q^{(s)}(\{P_i,Z_i\})\ ,
\ee
where
\be
\begin{array}{l}
 Q^{(1)}=H_{1,3}\,,\\
 Q^{(2)}=V_{1,23} V_{3,21} \,, \\
 Q^{(3)}=V_{1,23} V_{3,41}\,, \\
 Q^{(4)}=V_{1,43} V_{3,21} \,,\\
 Q^{(5)}=V_{1,43} V_{3,41} \,.
\end{array}
\ee
It is then straightforward to write the functions $g_{\l,s}$ in terms of the functions $F_s$,
\begin{align}
 g_1(r,\eta)&=\Acal \; f_4 F_5 \left(r^2-1\right)^2 \,,
 \nonumber\\
 g_2(r,\eta)&=\Acal \; f_1 \left(-2 r^2 \left(f_4 F_5+2 F_3 \eta  r\right)+f_2
   \left(2 F_4 \eta  r^2+F_2 \left(r^2+1\right)\right)-2 F_1
   \left(r^3+r\right)\right) ,
   \nonumber\\
 g_3(r,\eta)&=\Acal \; f_1 \left(2 f_4 F_5 r+2 \eta  \left(2 F_4 \eta 
   r^2+\left(r^2+1\right) \left(F_3 r^2+F_2\right)\right)+F_1
   \left(r^2+1\right)^2\right) ,
   \\
 g_4(r,\eta)&=\Acal \; f_1 \left(2 r \left(f_3 F_3+f_4 F_5\right)-f_2 \left(f_3
   F_4+2 F_2 r\right)+4 F_1 r^2\right) ,
   \nonumber\\
 g_5(r,\eta)&=\Acal \; f_1 \left(-f_3 \left(2 F_4 \eta +F_3
   \left(r^2+1\right)\right)-2 f_4 F_5-2 F_1 \left(r^3+r\right)-4
   F_2 \eta  r\right),
   \nonumber
\end{align}
where again we dropped the label  $\lambda$ and here the functions $f_i$ are given by
\begin{align}
\begin{split}
&f_1=r^2-2 \eta  r+1 \,,  \qquad \qquad \qquad  \qquad 
f_2=-r^2-2 \eta  r+1 \,,
\\ 
&f_3=-r^2+2 \eta  r+1\,,   \qquad \qquad \qquad  \qquad  
f_4=r^2+2 \eta  r+1\,,
\end{split}
\end{align}
and 
\be
\Acal(r,\eta)=\frac{2^{\sum_i\D_i}}{\left(1-r^2\right)^{2}\left(r^2+2 \eta  r+1\right)}
\left(\frac{r^2-2 \eta  r+1}{r^2+2 \eta  r+1}\right)^{\frac{1}{2} \left(\Delta _{12}-\Delta _{34}\right)}  \ .
\ee
Now we can already express the partial waves in terms of the function $W_s$. The last step is to obtain a recurrence relation for the coefficients $w_{s}$ of the series expansion (\ref{eq:seriesWpq}).

\begin{figure}
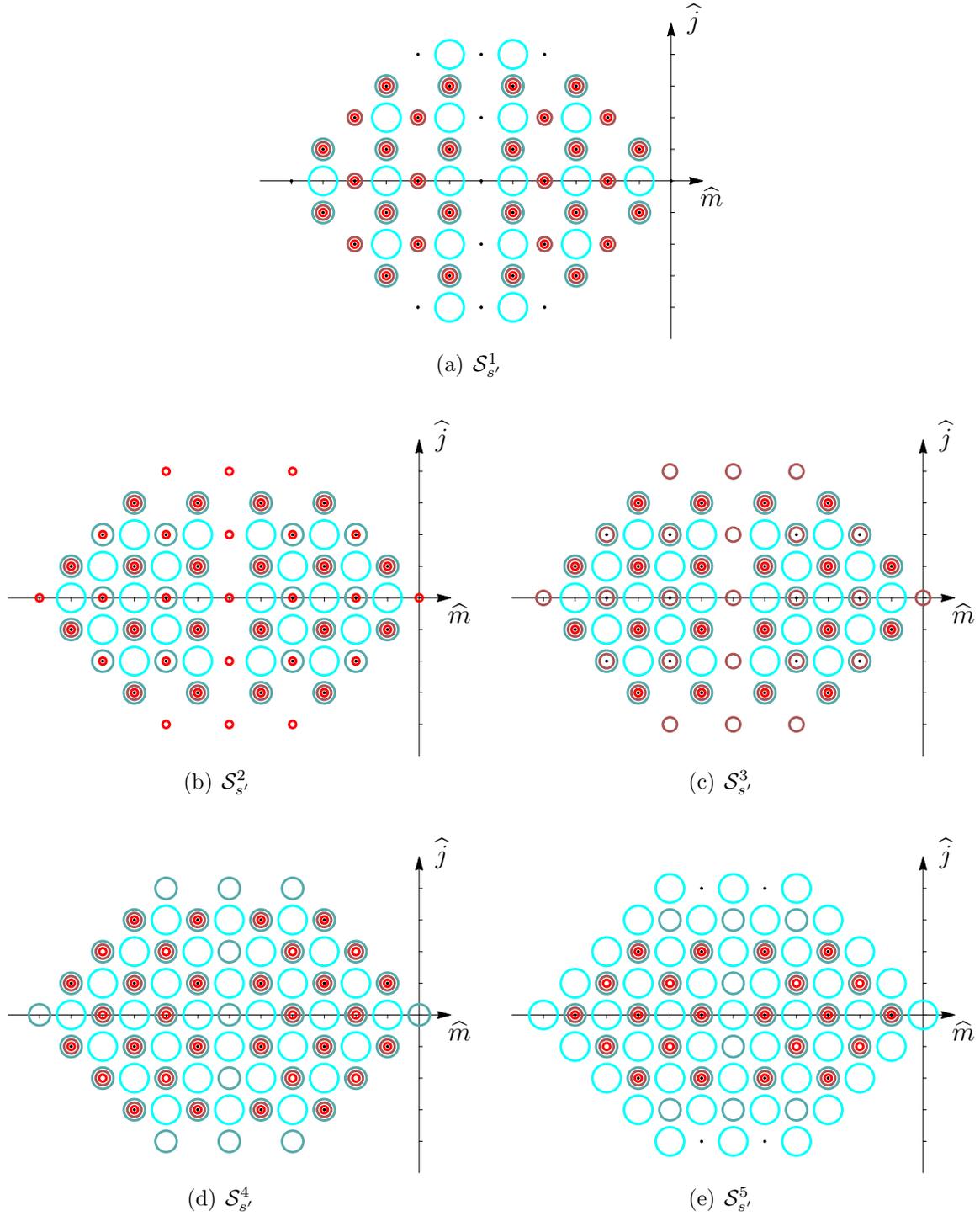

\centering
\subfigure[\label{fig:RecCoefficient2Vectors1} $\Scal^{1}_{s'}$ ]{\graphicspath{{Fig/}}
\def\svgwidth{7 cm} 
 \input{RecCoefficient2Vectors1.pdf_tex}}
\\
\subfigure[\label{fig:RecCoefficient2Vectors2} $\Scal^2_{s'}$ ]{\graphicspath{{Fig/}}
\def\svgwidth{7 cm} 
 \input{RecCoefficient2Vectors2.pdf_tex}}
~~~~
\subfigure[\label{fig:RecCoefficient2Vectors3} $\Scal^3_{s'}$ ]{\graphicspath{{Fig/}}
\def\svgwidth{7 cm} 
 \input{RecCoefficient2Vectors3.pdf_tex}}
\\
\subfigure[\label{fig:RecCoefficient2Vectors4} $\Scal^4_{s'}$]{\graphicspath{{Fig/}}
\def\svgwidth{7 cm} 
 \input{RecCoefficient2Vectors4.pdf_tex}}
~~~~
\subfigure[\label{fig:RecCoefficient2Vectors5} $\Scal^5_{s'}$ ]{\graphicspath{{Fig/}}
\def\svgwidth{7 cm} 
 \input{RecCoefficient2Vectors5.pdf_tex}}
\caption{\label{RecAVec2} Pictorial representation of the sets $\Scal^s_{s\rq{}}$ in formula (\ref{recrel2Vectors}). Increasing values of $s\rq{}$ correspond to increasing radius of the circles.
}
\end{figure}

As we did in the one external vector case we use the Casimir equation to obtain five coupled differential equations involving the functions $W_s$. Then we rewrite them in the basis of the expansion in such a way to get a recurrence relation for the coefficients (see appendix \ref{diffeqingegbasis}).
The result is a set of five coupled recurrence relations  
\be \label{recrel2Vectors}
\sum_{s'=1}^5 \sum_{(\hat{m},\hat{j}) \in \Scal^s_{s'}} c^s_{s'}(\hat{m},\hat{j}) \; w_{s'}(m+\hat{m},j+\hat{j})=0
 \ , \qquad s=1,\dots,5\ .
\ee
As before we represent the sets $ \Scal^s_{s\rq{}}$ in figure \ref{RecAVec2}. The colored circles are related to the value of $s\rq{}$ in (\ref{recrel2Vectors}), $s\rq{}=1$ corresponds to the black dot, increasing values of $s\rq{}$ correspond to larger circles. Each picture represents a different recurrence relation. 
The recurrence relations \eqref{recrel2Vectors} for $m=0$ reduce to 
\be
\begin{array}{r l}
[c_{(\D,j)}-c_{\lambda }]
 w_1(0,j) 
&=0 \,,\\
j [c_{(\D,j)}-c_{\lambda }]
 w_2(0,j)
&=0 \,,\\
 j [c_{(\D,j)}-c_{\lambda }]
w_3(0,j)
&=0 \,,\\
  j [c_{(\D,j)}-c_{\lambda }]
 w_4(0,j)
&=0 \,,\\
 j [c_{(\D,j,1)}-c_{\lambda }]
w_5(0,j)
&=0 \,.
\end{array}
\label{IC2spin1}
\ee
The coefficients multiplying $w_s(0,j)$ depend on the representation $\lambda$ of the exchanged primary operator. For the exchange of a traceless symmetric representation  $\lambda=(\Delta,l)$, it is easy to see that $w_s(0,j)=0$ for $j\neq l$. On the other hand, for $j=l$, equations \eqref{IC2spin1} only impose $w_5(0,l)=0$ leaving $w_s(0,l)$ for $s=1,2,3,4$ undetermined. These 4 independent solutions correspond precisely to the 4 independent conformal blocks that we parametrized by $(p,q)$ with $p,q=1,2$.
In the case of the representation $\lambda = (\Delta,l,1)$ we also find $w_s(0,j)=0$ for all $j\neq l$.
For $j=l$, equation  \eqref{IC2spin1} implies $w_s(0,l)=0$ for $s=1,2,3,4$, and leaves $w_5(0,l)$ as the only free parameter. This is consistent with the uniqueness of the conformal block $G_{\Delta ,l,1}$. 

It is interesting to notice that   equations
\eqref{IC2spin1} do not constrain 
$w_s(0,0)$ for $s=2,3,4,5$.
This should not come as a surprise since the  functions $W_s$ (with $s=2,3,4,5$) defined in (\ref{FtoW2vec}) appear with a derivative in $\eta$ which cancels the term $O(\eta^0)$. Therefore the only coefficients $w_{s}(j,m)$ (with $s=2,3,4 ,5$) that are physical are the ones with $j>0$.

The coefficients of the expansion of the functions $W_s$ have the same structure as shown in figure \ref{fig:Expansion_Plot}.
As an example we want to show the coefficients at level $m=1$ for $j=l+1$. In the case of the conformal blocks $G^{(p,q)}_{\D, l}$, these can be found inputting the initial conditions $w_1(0,l)= \d_{p,1}\d_{q,1}$, $w_2(0,l)= \d_{p,1}\d_{q,2}$, $w_3(0,l)= \d_{p,2}\d_{q,1}$, $w_4(0,l)= \d_{p,2}\d_{q,2}$, $w_5(0,l)= 0$ and using the recurrence relations \eqref{recrel2Vectors}. The result is
\begin{align}
\begin{split}
 w_1(1,l+1)=\frac{(2 h+l-2) \left(l \delta _{p,2}+\Delta _{12} \delta _{p,1}\right) \left(l \delta _{q,2}-\Delta _{34} \delta _{q,1}\right)}{(h+l-1) (\Delta +l)} \ , \\
 w_2(1,l+1)=\frac{(2 h+l-2) \left(l \delta _{p,2}+\Delta _{12} \delta _{p,1}\right) \left(\delta _{q,1}-\Delta _{34} l \delta _{q,2}\right)}{(l+1) (h+l-1) (\Delta +l)}  \ ,  \\
 w_3(1,l+1)=\frac{(2 h+l-2) \left(\Delta _{12} l \delta _{p,2}+\delta _{p,1}\right) \left(l \delta _{q,2}-\Delta _{34} \delta _{q,1}\right)}{(l+1) (h+l-1) (\Delta +l)}\ ,  \\
 w_4(1,l+1)=\frac{(2 h+l-2) \left(\Delta _{12} l \delta _{p,2}+\delta _{p,1}\right) \left(\delta _{q,1}-\Delta _{34} l \delta _{q,2}\right)}{(l+1)^2 (h+l-1) (\Delta +l)} \ , 
\end{split} 
\end{align}
and $w_5(1,l+1)=0$. Notice that the $w_5(m,j)$ do not vanish in general. For example, $w_5(1,l)\ne 0$ because there is a 
descendant of the primary ${\cal O}_{\D, l}$ in the representation $({\D+1, l,1})$ constructed by taking one derivative and antisymmetrizing it with an index of the operator. On the other hand, $w_5(n,l+n)= 0$  for any $n$, because the $n$-th symmetrized derivative constructs a descendant labeled by  $({\D+n, l+n})$. These features are also evident from figure \ref{RecAVec2}.

Similarly we can study the block $G_{\D,l,1}$ by setting the initial conditions $w_5(0,l)= 1$ and $w_s(0,l)=0$ for $s=1,2,3,4$. In this case we obtain 
\be 
 w_5(1,l+1)=-\frac{\Delta _{12} \Delta _{34} l (2 h+l-3)}{(l+1) (h+l-1) (\Delta +l)} \,,\\
\ee
and $w_s(1,l+1)=0$ for $s=1,2,3,4$. Again, by looking at  figure \ref{RecAVec2} one it is clear that $w_s(1,l) \neq 0$ (for $s=1,2,3,4$). Therefore all the five functions $W_s$ contribute to the construction of the conformal block at level $m>0$.

\subsection{Two external spin 2 operators and two scalars}
In this section we study all the conformal blocks appearing the decomposition of a four point function with two spin 2 tensors and two scalar operators,
\begin{align}
\begin{split}
&\langle \Ocal_1(P_1,Z_1)\Ocal_2(P_2) \Ocal_3(P_3,Z_3)\Ocal_4(P_4)\rangle =
  \sum_{\Ocal \in 
\ytableausetup{centertableaux,boxsize=0.3 em}
\begin{ytableau}
\null &&  & \\
\end{ytableau}
}
\; \sum_{p,q=3}^5 c^{(p)}_{12 \Ocal} c^{(q)}_{34 \Ocal} G^{(p,q)}_{\D, l}(P_i,Z_i)   \\
& + \sum_{\Ocal \in 
\ytableausetup{centertableaux,boxsize=0.3 em}
\begin{ytableau}
\null &&  & \\
\\
\end{ytableau}
}
\; \sum_{p,q=1}^2 c^{(p)}_{12 \Ocal} c^{(q)}_{34 \Ocal} G^{(p,q)}_{\D, l ,1}(P_i,Z_i)
+ \sum_{\Ocal \in 
\ytableausetup{centertableaux,boxsize=0.3 em}
\begin{ytableau}
\null &&  & \\
&\\
\end{ytableau}
}
c_{12 \Ocal} c_{34 \Ocal} G_{\D ,l , 2}(P_i,Z_i) \ .
\end{split}
\end{align}
There are now nine conformal partial waves $G^{(p,q)}_{\D, l}$ (for $p,q=3,4,5$) when $\Ocal$ is in the  symmetric and traceless representation $(\D,l)$, four $G^{(p,q)}_{\D, l ,1}$ (for $p,q=1,2$) if $\Ocal$ is in the representation $(\Delta ,l,1)$ and a single one $G_{\D, l ,2}$ when $\Ocal \in (\D,l,2)$. As in the two external vector case, we are going to study all the cases at once.

First we study the conformal block $\Gcal_{\l}$ on the cylinder  and  expand it in the allowed polynomial structures $p_{s}$ as described in \eqref{GcalToFGeneric}. 
In particular, there can exist $14$ possible structures 
\be
\begin{array}{clcl}
p_1&= \left(n\cdot z_1\right)^2 \left(n\cdot z_3\right)^2 \,,
& \qquad
p_8&= \left(n\cdot z_1\right) \left(n'\cdot z_1\right) \left(n'\cdot z_3\right)^2\,, \\
p_2&= \left(n\cdot z_1\right)^2 \left(n\cdot z_3\right) \left(n'\cdot z_3\right),
  & \qquad
p_9&=\left(n'\cdot z_1\right)^2 \left(n'\cdot z_3\right)^2 \,,\\
p_3&= \left(n\cdot z_1\right) \left(n\cdot z_3\right)^2 \left(n'\cdot z_1\right),
  & \qquad
p_{10}&= \left(z_1\cdot z_3\right) \left(n\cdot z_1\right) \left(n\cdot z_3\right), \\
p_4&= \left(n\cdot z_1\right)^2 \left(n'\cdot z_3\right)^2 \,,
& \qquad
p_{11}&= \left(z_1\cdot z_3\right) \left(n'\cdot z_1\right) \left(n\cdot z_3\right) ,   \\
 p_5&=\left(n\cdot z_1\right) \left(n'\cdot z_1\right) \left(n\cdot z_3\right)
   \left(n'\cdot z_3\right),
& \qquad
p_{12}&=\left(z_1\cdot z_3\right) \left(n\cdot z_1\right)  \left(n'\cdot z_3\right),
   \\
p_6&=\left(n'\cdot z_1\right)^2 \left(n\cdot z_3\right)^2 \,,
& \qquad
p_{13}&= \left(z_1\cdot z_3\right) \left(n'\cdot z_1\right) \left(n'\cdot z_3\right),
   \\
p_7&= \left(n'\cdot z_1\right)^2 \left(n'\cdot z_3\right) \left(n\cdot z_3\right) ,
 & \qquad
p_{14}&= \left(z_1\cdot z_3\right)^2 \,,
\end{array} 
\ee
therefore 
\be
\label{FStructures2StressTensors}
\Gcal_{\l}= \sum_{s=1}^{14} F_s(r,\eta)\, p_s \ .
\ee
To write a nice expansion for the functions $F_s$, we study the decomposition  of the four point function in eigenstates of the cylinder Hamiltonian. We obtain
\begin{align}
\begin{split}
\Gcal_{\l}= &
\sum_{m=0}^\infty r^{\Delta+m} 
\sum_j  \sum_a
\Big[N_\mathcal{T}^2 \;  u_0 \tilde u_0\; (z_1)_{\mu_1}(z_1)_{\mu_2} (z_3)_{\nu_1}(z_3)_{\nu_2} \mathcal{T}_j^{\mu_1 \mu_2 \nu_1\nu_2}(n,n')
\label{SphericalHarmonicExpansion2}
\\
&
\quad +
N_\mathcal{V}^2 \;
\big( 
u_1\,z_1\cdot n +
u_2\, z_1\cdot \nabla_n 
\big)\big( 
\tilde u_1 \,z_3\cdot n' +
\tilde u_2\, z_3\cdot \nabla_{n'} 
\big)  (z_1)_\mu (z_3)_\nu \mathcal{V}_j^{\mu\nu}(n,n') 
 \\
&
\qquad\qquad+
\big( 
u_3\,(z_1\cdot n)^2+
u_4\,z_1\cdot n \, z_1\cdot \nabla_n +
u_5\,(z_1\cdot \nabla_n)^2 
\big)
 \\
&
\qquad\qquad\qquad
\times \big( 
\tilde u_3\,(z_3\cdot n')^2 +
\tilde u_4 \,z_3\cdot n' \, z_3\cdot \nabla_{n'} +
\tilde u_5\,(z_3\cdot \nabla_{n'})^2
\big) \mathcal{C}_j(n\cdot n') \Big]\,,
\end{split}
\end{align}
where $u_s$ and $\tilde u_s$ are coefficients dependent on $j$, $m$ and $a$. As in the two external vectors case, we chose to factorize out the coefficients  $N_\mathcal{V}^2$ and $N_\mathcal{T}^2$ in order to obtain simpler formulae.
 Again one has to fix their value at $m=0$ depending on the possible primary representation $\l$ that is exchanged in the conformal block. The tensor harmonic
\ytableausetup{centertableaux,boxsize=1.2 em}
\begin{align}
\mathcal{T}_j^{\mu_1 \mu_2 \nu_1\nu_2}(n,n')&=
 n_{ \a_1}\dots n_{\a_j}
\pi\Big(
{ \scriptsize
\begin{ytableau}
\a_1&\a_2&\, _{\cdots}&\a_j \\
\mu_1 &\mu_2
\end{ytableau}
}
,
{ \scriptsize
\begin{ytableau}
\b_1&\b_2&\, _{\cdots}&\b_j \\
\nu_1&\nu_2 
\end{ytableau}
}
\Big) 
 n'_{ \b_1}\dots n'_{\b_j}
\end{align} 
can be written in terms of scalar harmonics as shown  in \cite{projectors}.
From this representation it is straightforward to find a set of $14$ functions that possess a nice radial expansion, which we can label as follows 
\be
W_{p, q}= \sum_{j,m} w_{p, q}(j,m) r^{\D+m} \mathcal{C}_j(\eta) \,, \qquad\qquad
\left\{
\begin{array}{l}
p=q=0 \\
p,q \in \{1,2\} \\
p,q \in \{3,4,5\}  \\
\end{array}
\right. \ ,
\ee
where
\be
w_{p, q}(j,m) =\sum_a u_p(m,j,a) \tilde u_q(m,j,a) \ .
\ee
Next we want to relate the functions $W_{p,q}$ to the conformal partial waves.
As usual the first step is to find a relation between $W_{p,q}$ and $F_s$ according to \eqref{FStructures2StressTensors}.
The result is, for example,\footnote{
For the sake of brevity  the remaining thirteen equations are appended in a \Mathematica file.}
\begin{align}
\begin{split}
F_1=&\partial _{\eta }{}^2W_{3,5}-2 \partial _{\eta }{}^2W_{4,5}+6 \partial
   _{\eta }{}^2W_{5,5}+\partial _{\eta }{}^3W_{1,2}-3 \partial _{\eta}{}^3W_{2,2}+\big(3+4 (-2+h) h \big) \partial _{\eta }{}^4W_{0,0} \\
&+\eta 
   \left(-\partial _{\eta }{}^3W_{4,5}+6 \partial _{\eta
   }{}^3W_{5,5}-\partial _{\eta }{}^4W_{2,2}+\eta  \partial _{\eta
   }{}^4W_{5,5}\right) .
\end{split}
\end{align}
We then write the conformal partial wave in terms of the conformal blocks $g_s$ using \eqref{GTog_Generic}, where now we have the following allowed structures 
\be
\begin{array}{c l c c l}
 Q_1&=H_{1,3}^2 \,,
& \qquad&
 Q_8&=H_{1,3} V_{1,4,3} V_{3,2,1} \,,\\
 Q_2&=V_{1,2,3}^2 V_{3,2,1}^2 \,,
& \qquad&
 Q_9&=H_{1,3} V_{1,4,3} V_{3,4,1} \,,\\
 Q_3&=V_{1,2,3}^2 V_{3,4,1}^2 \,,
& \qquad&
 Q_{10}&=V_{1,2,3} V_{1,4,3} V_{3,2,1}^2 \,,\\
 Q_4&=V_{1,4,3}^2 V_{3,2,1}^2 \,,
& \qquad&
 Q_{11}&=V_{1,2,3} V_{1,4,3} V_{3,4,1}^2 \,,\\
 Q_5&=V_{1,4,3}^2 V_{3,4,1}^2 \,,
& \qquad&
 Q_{12}&=V_{1,2,3}^2 V_{3,2,1} V_{3,4,1} \,,\\
 Q_6&=H_{1,3} V_{1,2,3} V_{3,2,1} \,,
& \qquad&
 Q_{13}&=V_{1,4,3}^2 V_{3,2,1} V_{3,4,1} \,,\\
 Q_7&=H_{1,3} V_{1,2,3} V_{3,4,1} \,,
& \qquad&
 Q_{14}&=V_{1,2,3} V_{1,4,3} V_{3,2,1} V_{3,4,1}\,, \\
\end{array}
\ee
Therefore we obtain linear relations between the $F_s$ and $g_s$, for example,\footnote{
Again the full set of relations is presented in a  \Mathematica file.}
\be
g_1= 
2^{\sum_i \D_i} \left(\frac{r^2-2 \eta  r+1}{r^2+2 \eta  r+1}\right)^{\frac{1}{2} \left(\Delta _{12}-\Delta _{34}\right)} 
F_{14}\,.
\ee

The result of these redefinitions  is that the Casimir equation can be written as a set of 14 coupled differential equations on the functions $W_{p,q}$, which we present in a  \Mathematica file. It is then a tedious exercise to rewrite them as fourteen algebraic relations for the coefficients  $w_{p,q}$, as explained in the previous sections. We leave this exercise for the reader. 

There are three allowed sets of initial conditions for the coefficients   $w_{p,q}$. When the exchanged representation is labeled by $\l=(\D,l)$  there exist nine conformal blocks $G_{\D,l}^{(p,q)}$ with  $p,q\in \{3,4,5\}$. We can obtain each block $G_{\D,l}^{(p,q)}$ by setting as an initial condition the function $w_{p,q}(0,j)$ (with the same $p$ and $q$) to $\d_{j,l}$ and all the other $w_{p\rq{},q\rq{}}(0,j)$ to zero. In other words, for a given  $p,q\in \{3,4,5\}$,  the set of fourteen $w_{p',q'}(0,j)$  are defined by the initial
condition
\be
w_{p\rq{},q\rq{}}(0,j)= \d_{p\rq{},p}\d_{q\rq{},q} \d_{j,l} \,.
\ee
Similarly, when the exchanged representation is  $\l=(\D,l,1)$, we have four blocks $G_{\D,l,1}^{(p,q)}$ with  $p,q\in \{1,2\}$ which can be obtained by setting 
\be
w_{p\rq{},q\rq{}}(0,j)= \d_{p\rq{},p}\d_{q\rq{},q} \d_{j,l} \,.
\ee
Finally, when we exchange $\l=(\D,l,2)$, there exists a single block $G_{\D,l,2}$ which is obtained by setting 
\be
w_{p\rq{},q\rq{}}(0,j)=\d_{p\rq{},0}\d_{q\rq{},0} \d_{j,l} \,.
\ee
We checked that these choices of initial condition actually solve the corresponding Casimir equation at lowest order in $r$ and therefore reproduce the correct leading OPE behavior of the four-point function.

\section{Recursion relation from analytic structure in $\D$}
\label{RecursionInDelta}
In this section we want to use the alternative strategy of \cite{ arXiv:1509.00428} in order to obtain a radial expansion for
 all the conformal blocks that can appear in the conformal block decomposition 
\eqref{eq:4ptCBdecomposition}
for the four-point function of two vector and two scalar operators.

Following the idea of \cite{arXiv:1307.6856}, in \cite{arXiv:1509.00428} a different method 
to obtain the $r$-series expansion of 
  the conformal blocks is described.
  The main strategy is to find recurrence relations from the analytic structure of the conformal blocks in the conformal dimension $\D$ of the exchanged operator. 
In \cite{arXiv:1509.00428} the full set of poles $\D^\star_A$ of a bosonic conformal block $G_\l^{(p,q)}$ for the exchange the most generic representation $\l=(\D,l_1,\dots,l_{[h]})$ was obtained. Moreover, it was explained that the residue at a pole $\D^\star_A$ is just a linear combination of new conformal blocks $G_{\l_A}^{(p\rq{},q\rq{})}$ associated to the exchange of a (null) primary descendant with   weight given by $\l_A=(\D_A,l_{1 A},\dots,l_{[h] A})$, namely
\be
G_\l^{(p,q)} \sim \frac{1}{\D-\D^\star_A }\sum_{p\rq{},q\rq{}} (R_A)_{p p\rq{} q q\rq{}} G_{\l_A}^{(p\rq{},q\rq{})} \ .
\ee
It was also explained how to obtain the coefficients $(R_A)_{p p\rq{} q q\rq{}}$ from a direct computation. For concreteness we write the full set of poles $\D^\star_A$ and associated primary descendants $\l_A$ in the following table
\be \label{AllTheLabels}
\begin{array}{|l | ccc|}
\hline
\phantom{\Big(}	\qquad \quad\quad	A								&\D^\star_A 					&n_A 										&l_{k A}	 \\ 
\hline
\phantom{\Big(}  \I_k, \; \;n: n\in [1,l_{k-1}-l_k]				& k-l_k-n 			 		& n 	  	   		&\quad l_k + n \quad \\ 
\phantom{\Big(} \II_k,  \;n: n\in [1,l_{k}-l_{k+1}\,]  \;\; 	&      \quad   2h+l_k-k-n \quad & \quad n \quad  			&\quad l_k - n \quad  \\
\phantom{\Big(} \III,     \, \,   n: n \in [1,\infty) \; 				& h-n  						& 2n       			 	 &l_k     \\ 
\phantom{\Big(} \IV,     \, \,   n: n \in [1,l_{[h]}] \; 				& h+\frac{1}{2}-n  					& 2n-1       			 	 &l_k     \\ 
\hline
\end{array}
\ee
where $k=1,\dots ,[h]$ and $n$ is an integer. To obtain $\l_A$ from the data of the table it is enough to know that $\D_A=\D^\star_A+n_A$, 
and that all the $l_{i A}$ which are not represented in the table are left unchanged ($l_{i A}=l_i$ for $i \neq k$). 
 We can then reconstruct the full conformal block summing over all the poles in $\D$ and over the regular part.
In radial coordinates this can be done by writing 
\be
g^{(p,q)}_{\l,s}(r,\eta)=(4r)^\D h^{(p,q)}_{\l,s}(r,\eta) \,,
\ee
where 
\be
h^{(p,q)}_{\l,s}(r,\eta)=
h^{(p,q)}_{\infty,\l ,s}(r,\eta)+ \sum_A\sum_{p\rq{},q\rq{}}  (4r)^{n_A} \frac{(R_A)_{pp\rq{} q q\rq{}}}{\D-\D^\star_A}  h^{(p\rq{},q\rq{})}_{\l_A,s}(r,\eta) \,,
\label{eq:recinDelta}
\ee
and the regular part $h^{(p,q)}_{\infty,\l ,s}(r,\eta)= \lim_{\D \rightarrow \infty } h^{(p,q)}_{\l,s}(r,\eta)$ can be computed solving the Casimir equation at  leading order in large $\D$, as explained in  \cite{ arXiv:1509.00428}. 
Given the knowledge of the poles $\D_A^\star$, the coefficients $R_A$ and the conformal block at infinity $h_\infty$, one can use \eqref{eq:recinDelta} to obtain a series expansion (in $r$) for the conformal block.
This is usually more efficient than the method of the previous section. 

\subsection{Recursion relation for two external vectors and two scalars}
We are now ready to apply the method  of \cite{ arXiv:1509.00428} to compute all the conformal blocks appearing in the four point function of two external vectors.

Let us start by studying the conformal block $g_{\l,s}^{(p,q)}$ for the exchange of a symmetric and traceless representation $\l=(\D,l)$. The poles $\D^\star_A$ and the residue $g_{\l_A,s}^{(p,q)}$ which can couple to the external states are characterized by the following table
\be
\begin{array}{|l | ccc c|}
\hline
\phantom{\Big(}	\qquad \quad\quad	A														&\D^\star_A 					&n_A 										&l_{1A}	&l_{2A}	 \\ 
\hline
\phantom{\Big(}  \I_1, \; \;n: n\in [1,\infty)				& 1-l-n 			 				& n 	  	   						 &l + n& \quad 0 \quad  \\ 
\phantom{\Big(} \II_1,  \;n: n\in [1,l \,]  \;\; 	&\quad   l+2h-1-n  	\quad & \quad n \quad  						&\quad l - n \quad  & \quad 0 \quad\\
\phantom{\Big(} \III,     \, \,   n: n \in [1,\infty) \; 				& h-n  							& 2n       			 						 &l    & \quad 0 \quad   \\ 
\phantom{\Big(} \I_2,   \;\; n: n=1 \; 				& 1 							&1      			 						 &l     & \quad 1 \quad  \\ 
\hline
\end{array}
\ee
Therefore we can write the following recurrence relation 
\begin{align}
\label{eq:rechsymtraceless}
\begin{split}
h^{(p,q)}_{(\D,l),s}(r,\eta)=\ 
&h^{(p,q)}_{(\infty,l),s}(r,\eta)+ \sum_{A }\sum_{p\rq{},q\rq{}=1}^2 (4r)^{n_A} \frac{(R_A)_{pp\rq{} q q\rq{}}}{\D-\D^\star_A}  h^{(p\rq{},q\rq{})}_{(\Delta_A,l_A),s}(r,\eta) \\
&+ \frac{(4 r)}{\D-1} (R_{\I_2 , 1})_{p q} \;  h_{(2, l,1 ),s}(r,\eta)\,,
\end{split}
\end{align}
where the sum in $A$ runs over $(\I_1,n), (\II_1,n)$ and $(\III,n)$. The coefficients $(R_A)_{pp\rq{} q q\rq{}}$ are given in appendix \ref{appendix:recdelta} using the results of \cite{ arXiv:1509.00428}.
In appendix \ref{appendix:recdelta} we also explain how to find $(R_{\I_2 , 1})_{p q}$, which was not previously computed, \footnote{Here the notation $(a,b)_p$ means that we pick the component $p$ of the vector $(a,b)$.}
\be
\label{RI21}
\big(R_{\I_2 , 1}\big)_{p q}=\frac{1}{2 (l+1) (2 h+l-3)} \;\;  \Big( 1\, , \; - \D_{12}  \, \Big)_p \;\; \Big(1\,, \;   \D_{34} \, \Big)_q\ .
\ee
In appendix \ref{ConformalBlockAtLargeDelta} we also compute $h^{(p,q)}_{(\infty,l),s}(r,\eta)$ following the recipe proposed in \cite{ arXiv:1509.00428}.

When we exchange a representation $\l=(\D,l,1)$ we have the following set of poles
\be
\begin{array}{|l | ccc c|}
\hline
\phantom{\Big(}	\qquad \quad\quad	A														&\D^\star_A 					&n_A 										&l_{1A}	&l_{2A}	 \\ 
\hline
\phantom{\Big(}  \I_1, \; \;n: n\in [1,\infty)				& 1-l-n 			 				& n 	  	   						 &l + n& \quad 1 \quad  \\ 
\phantom{\Big(} \II_1,  \;n: n\in [1,l -1 \,]  \;\;	&\quad   l+2h-1-n  	\quad & \quad n \quad  						&\quad l - n \quad  & \quad 1 \quad\\
\phantom{\Big(} \III,     \, \,   n: n \in [1,\infty) \; 				& h-n  							& 2n       			 						 &l    & \quad 1 \quad   \\ 
\phantom{\Big(} \II_2,   \; n: n=1 \; 				&  2h-2 							&1      			 						 &l     & \quad 0 \quad  \\ 
\hline
\end{array}
\ee
We can write the following recurrence relation to determine the block $h_{(\D,l,1)}$
\begin{align}
\label{recrelh00}
\begin{split}
h_{(\D,l,1),s}(r,\eta)=\ 
&h_{(\infty,l,1),s}(r,\eta)+ \sum_A (4r)^{n_A} \frac{R_A}{\D-\D^\star_A} h_{(\Delta_A,l_A,1),s}(r,\eta) \\
&+ (4r)  \sum_{p,q=1}^2 \frac{(R_{\II_2,1})_{p q} }{\D-(d-2)} \;  h^{(p,q)}_{(d-1, l),s}(r,\eta) \ ,
\end{split}
\end{align}
where again $\sum_A$ runs over $(\I_1,n), (\II_1,n)$ and $(\III,n)$. Since $R_{A}$ are just numbers (not matrices), they can be easily obtained matching the coefficients from the expansion in $r$. We present them in appendix \ref{appendix:recdelta}.
Moreover in appendix \ref{appendix:recdelta} we also explain how to compute $R_{\II_2,1}$ from first principles,
\be
\label{RII21}
\big(R_{\II_2,1}\big)_{p q}=\frac{l (h-1) }{2 h+l-2} \left(2 h+l-2, \; \frac{\Delta _{12}}{l}\right)_{p} \left(2 h+l-2, \; -\frac{\Delta _{34}}{l}\right)_q \ .
\ee
Finally in appendix \eqref{ConformalBlockAtLargeDelta} we also find $h_{(\infty,l,1),s}(r,\eta)$.

A collection of all the definitions needed to obtain the recursion relation for the blocks is presented in a  \Mathematica file included with the submission.

We would like to comment on an important feature of the recursion relation \eqref{eq:recinDelta}.
At first sight, the reader might be puzzled by the fact that for most of the poles $\D^\star_A$ of formula \eqref{eq:recinDelta} the residue looks divergent because the conformal blocks $ G_{\l_A}^{(p\rq{},q\rq{})}$ have a pole precisely at this value of $\D$, for every $A$ which is not of the type $(\II_1,n)$ and $(\III,n)$. 
In the example above, this phenomena happens in (\ref{recrelh00})  for the residue of the block $h_{(\D,l,1),s}$ at the pole $\D=(d-2)$. In fact, the term
\be
\label{problematicpoleII21}
 \sum_{p,q=1}^2 (R_{\II_2,1})_{p q}  \;  h^{(p,q)}_{(d-1, l),s}(r,\eta)\,,
\ee
in the second line of (\ref{recrelh00}), looks divergent because equation \eqref{eq:rechsymtraceless} tells us that $h^{(p,q)}_{(\D, l),s}(r,\eta)$ has a pole at $\D=d-1$.
However, one can easily check that $R_{\II_2,1}$, as defined in (\ref{RII21}), is such that the combination (\ref{problematicpoleII21}) is actually finite.
We expect this to happen in general so that, in odd spacetime dimension, the only singularities of the conformal blocks in the $\D$--plane are simple poles. Moreover, this analytic structure should extend to non-integer spacetime dimension  by analytic continuation. \footnote{In even spacetime dimension there are higher order poles in the $\D$--plane.}

In this section, we presented the recursion relations \eqref{eq:rechsymtraceless} and (\ref{recrelh00}) which allow us to determine all the conformal blocks that can appear in the conformal block decomposition 
\eqref{eq:4ptCBdecomposition}
of the four-point function with two vector and two scalar operators. 
In \cite{arXiv:1109.6321}, it was shown that the conformal blocks associated to the exchange of symmetric traceless primary operators could be written as differential operators acting on scalar conformal blocks.
This gives an alternative way to determine $ h^{(p,q)}_{(\D, l),s}(r,\eta)$ (see appendix \ref{Spinning_CB}).
On the other hand,  $ h_{(\D, l,1),s}(r,\eta)$ must be determined using (\ref{recrelh00}). Notice that in this context, the last term in  (\ref{recrelh00}) is a source term  that can be written in terms of (derivatives of) the scalar conformal blocks.  
This is a specific realization of the strategy purposed in \cite{arXiv:1109.6321, arXiv:1505.03750}. They suggested that one should start by computing \seed conformal blocks, which have the simplest external operators that can exchange a primary in a given $SO(d)$ irreducible representation. For example,  the  \seed blocks for symmetric traceless operators are scalar CBs.
Then, one can create external spin by acting with appropriate differential operators \cite{arXiv:1109.6321, arXiv:1505.03750,  arXiv:1508.00012}.
The \seed conformal blocks have the advantage that they are unique, in the sense that there is only one block for the chosen external operators.
This means that the recursion relations \eqref{eq:recinDelta} are simpler for \seed CBs.
This is clear in the particular example of equation (\ref{recrelh00}). This is just one equation, where the poles are proportional to the same CB with the exception of one pole that can be determined from other \seed CBs.
\section{Conclusion}
In the first part of this paper we generalized the series expansion in the radial coordinate introduced by \cite{Hogervorst:2013sma} for general CBs in any spacetime dimension. To do so we defined an expansion which inherits the structure of the conformal representations exchanged in the CB. We considered the CB as 
a projection of a four-point function obtained by summing over a complete basis of a conformal representation. We organized the sum by grouping together all the descendant operators in the same irreducible representations ${\bf  Y}$ of $SO(d)$. 
Each state gives a contribution proportional to $r^E$ where $E$ is the eigenvalue of the cylinder hamiltonian.  
Therefore, we naturally obtained the CB in an expansion in radial coordinates.
We implemented the method in order to build  CBs when the external operators are in the traceless and symmetric representation, but one could generalize it for external fermionic operators or for mixed symmetry external operators  \cite{arXiv:1411.7351}. We exemplified the method in four cases, for the scalar CB, for one external vector and three scalars, two vectors and two scalars and finally for two spin two operators and two scalars.
 
Another useful result of this work is to explain how to write closed form recurrence relations for the coefficients of the expansion of the functions  $W_s(r,\eta)$ that define the CB. These can be obtained by following an algorithm, which we give. On the other hand there can exist infinitely many equivalent recurrence relations and it is less straightforward to find efficient ones. We give recipes to simplify them, however it is still not clear which are the optimal ones and how to find them. We give examples of such simplified recurrence relations in three cases: the scalar CB, one external vector and three scalars, and two vectors and two scalars. 

One of the nice features of these relations is that they are analytical results, therefore they can be used to extract many interesting properties of the coefficients of the expansions.
For example one can consider explicitly  the limit $\D \rightarrow \infty$ of the recurrence relations and check that the result is finite.
Also, for the scalar block one can trivially check that for $\D_{12}=0=\D_{34}$, all the coefficients at odd level are identically zero. For the other CBs the condition $\D_{12}=0=\D_{34}$ is even more interesting:
for one external vector and three scalars the two coupled recurrence relations completely decouple, while in the two external vectors case, the five relations form two decoupled groups. Other limit cases, like large $l$ and large $h$, can be also approached analytically using the recurrence relations. 
 The reader is encouraged to play with the \Mathematica files in these respects.

The algorithm to build CBs from the recurrence relations is quite efficient for numerical evaluations. However when many parameters (like $h$, $\D_{12}$, $\D_{34}$, $\D$, $l$) are left unfixed, the recurrence relations may become slow. This happens because every  coefficient at a certain level is written in terms of a large linear combination of coefficients at lower levels, which are very non trivial rational functions of the parameters. Therefore it may take a long time to simplify the expressions. This problem of course does not subsist when the parameters are numbers, since the simplifications become trivial.

This method has also the feature of generating at the same time the span of all the CBs which can be exchanged in a given four point function. In fact, to compute only one CB, one needs to set to zero all the initial conditions correspondent to the other ones. Because of this feature, it is not straightforward to take advantage of the philosophy of the \seed blocks, for which one should study the simplest blocks for a given exchange, and then act with derivatives to obtain more complicated ones. In fact, the relation between the functions $W_s$ and the CBs is not trivial since each CB is written as a linear combination of derivatives of $W_s$. Therefore the relation between the functions $W_s$ of a given CB and the ones of its \seed block takes the form of a complicated differential equation. However, it may be  useful to analyze this problem in more detail to find an efficient strategy to include the information about the \seed blocks inside the expansion. 

The importance of the \seed strategy can be easily understood when we consider more complicated four point functions. For example, to study four different spin 2 operators one needs to find the expansion of $633$ functions $W_s$ which can be obtained from the same number of coupled recurrence relations. Even if the technology to do this is ready at the present day (due to the projectors defined in \cite{projectors}), this strategy  is clearly going to be very inefficient. On the other hand the $633$ CBs can be obtained from appropriate derivatives of just nine \seed blocks which one may derive using a combination of these and other techniques.

It would also be important to find a way of implementing the conservation of external operators in order to decrease the number of functions $W_s$ to compute. In fact, the most interesting CBs for numerical bootstrap studies are the ones of four conserved operators, in particular vector currents or stress tensors. In these cases the number of allowed CBs reduces drastically: in general dimensions there are only 7 CBs for four equal vector currents and 29 for four stress tensors \cite{arXiv:1311.4546}. If one could define a single function $W_s$ for each CB, the resulting problem would look accessible (surely for the four vector currents case) to the radial expansion even without any further technical development.

In this paper we also applied the alternative
strategy of \cite{arXiv:1509.00428}, which as well builds the CBs from their radial expansion. The main idea of this method is to  study the behavior of CBs in the complex $\D$ plane \cite{Zamolodchikov:1985ie,arXiv:1307.6856}. From the analysis of \cite{arXiv:1509.00428} we know that the CBs  are meromorphic functions of $\D$ (in odd dimensions $d$, but it is possible to analytically continue the method to any non even $d$), and that the residues at all the poles can always be written as linear combination of other  CBs labeled by the position of primary descendants which can appear in the conformal representation. This fact naturally leads to recursion relations.
 All the poles and the CBs at each residue are known in general while the specific linear combination has to be computed case by case following a systematized algorithm. The regular part of the CBs needs also to be computed case by case  and it can be thought  as the initial condition for the recursion relation.\footnote{We believe that also this step can be done in quite  general cases; we actually already have this result for four external vectors CBs.} The only important subtlety, which was noted in \cite{arXiv:1509.00428}, is that the CBs which appear in the residues are actually divergent. This problem was preventing the computation of new CBs using this technology.  Here we checked in one explicit case that the divergent CBs at the residue actually appear in a very special linear combination which exactly cancels the divergent piece,  giving a finite result. We  believe that this should always happen and that this method can be safely used to build any CB.

In particular in this paper we obtained all the five CBs in the scalar-vector-scalar-vector four point function: four of them correspond to the exchange of a symmetric and traceless representation, the fifth is a \seed block for the exchange of an operator  $(\D,l,1)$. The  recurrence relation which determines the \seed block couples to the other four CBs, but in a linear way. Thus one can simply replace the four CBs by derivatives of the scalar CB (which is the \seed block for the exchange of a traceless and symmetric operator). 
This is an explicit realization of the \seed block program.
Moreover this structure generalizes for any CB since the recursion relations of \cite{arXiv:1509.00428} always relate the CBs in a linear way. Therefore in the recurrence relations to determine any \seed CB one will only need to impute derivatives of simpler \seed blocks, recursively up to the scalar CB. In addition, as discussed in \cite{arXiv:1509.00428}, the conservation of external operators can be directly incorporated in this method in order to simplify the equations. And finally, this method allows a more efficient computation of CBs even in the case of many unfixed parameters  (like $h$, $\D_{12}$, $\D_{34}$, $\D$, $l$). In fact, at each step of these recurrence relations all the coefficients are known in a closed form, so there is no need to simplify them.

The only downside of this method is that it can be tedious  to perform all the computations needed to find the closed form formula for the linear combination of CBs at each pole. Sometimes it may help to have another way to compute the CB and use it to determine some of these coefficients. In fact this is what we did in the two external vectors case, since we already knew a simpler way to build the CBs from their radial expansion. We therefore believe that the interplay of these two methods can be very profitable in the plan of obtaining spinning conformal blocks.

\section*{Acknowledgements}
This research received funding from the [European Union] 7th Framework Programme (Marie Curie Actions) under grant agreements No 269217 and 317089 (GATIS), and from the research grant CERN/FIS-NUC/0045/2015.
The work of E.T. has been supported by the Portuguese Fundac\~ao para a Ci\^encia e a Tecnologia (FCT) through the fellowship SFRH/BD/51984/2012. His research was partially supported by Perimeter Institute for Theoretical Physics. Research at Perimeter Institute is supported by the Government of Canada through Industry Canada and by the Province of Ontario though the Ministry of Economic Development \& innovation.


\appendix 
\section{Differential equations in Gegenbauer basis}
\label{diffeqingegbasis}
In the main text we explain how to trade the differential equations for the functions $W_s$, which come from the action of the conformal Casimir, with algebraic relations for the coefficients $w_s$ of the expansion of the functions $W_s$.
However, we did not stress that for one  differential equation one can in principle write infinite algebraic relations to which would correspond different \lq\lq{}dot diagrams\rq\rq{} like the ones depicted in figures \ref{fig:RecCoefficient1Vector1} and \ref{fig:RecCoefficient1Vector2}. This is not surprising since given a relation as for example \eqref{recrelScalar}, it is alway possible to sum to it the same relation with shifted $m$ and $j$. The result can still be used as a valid recurrence relation for the coefficients, but it would in general contain more terms and therefore it would be less efficient.
The aim of this appendix is to explain the main manipulations which we used to simplify the algebraic  relation for the $w_s$.

The first caveat is that in general, instead of the last relation of (\ref{action_r_eta}), it is better to use 
\begin{align}
\begin{split}
\partial^{n}_\eta \mathfrak f_{m,l}(r,\eta)= \frac{1}{\eta ^2-1}\Big[
&
 (5-2 h-2 n)\, \eta \, \partial^{n-1}_\eta \mathfrak f_{m,l}(r,\eta )
\\
&
+(l-n+2) (2 h+l+n-4) \partial^{n-2}_\eta \mathfrak f_{m,l}(r,\eta )
 \Big]\ ,
\qquad (n>1)
\end{split}
\label{eta_derivatives}
\end{align}
 to express any derivative in $\eta$ of order higher than one. Formula \eqref{eta_derivatives} comes from the usual Gegenbauer differential equation, once we derived it $n-2$ times.
This is enough to obtain formula \eqref{recrelScalar} for the scalar conformal block.

For more general cases the procedure is more complicated.
In fact, in the scalar case the Casimir equation gives just the single differential equation \eqref{ScalarCasimirEq}, while  for any other case we obtain a system of coupled differential equations that we schematically write as  $\texttt{PDE}_s=0$ for the set of functions $W_s$, where the label $s$ goes from $1$ to the total number of tensor structures in the four-point function. Notice that
each equation $\texttt{PDE}_s=0$  arises by collecting the terms multiplying the tensor structure $Q^{(s)}$ in the Casimir equation.
However, the choice of the basis $Q^{(s)}$ is not unique. 

\begin{figure}
\centering
{\graphicspath{{Fig/}}
\def\svgwidth{5 cm} 
 \input{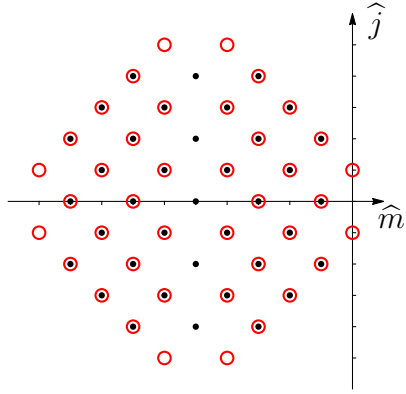}}
\caption{ \label{fig:RecCoefficient1Vector2WRONG}The set $\tilde \Scal_{s}$ appearing in the linear combination (\ref{recrel1VectorWRONG}).
}
\end{figure}

For example, in the case of one external vector the Casimir equation gives rise to a system of two coupled differential equations for the two functions $W_1$ and $W_2$, that we schematically write as
\be
\texttt{PDE}_1(W_1,W_2)=0\,, \qquad \qquad
\texttt{PDE}_2(W_1,W_2)=0 
\,.
\ee
We could write the differential equations $\texttt{PDE}_s$ explicitly, but the expressions are very lengthy and they do not provide any deep insight. We want however to stress that both  $\texttt{PDE}_1$ and  $\texttt{PDE}_2$ can be written as a combination of the terms
\be
\left\{W_1,\partial _{\eta }W_1,\partial _{\eta }W_2,\partial _{\eta
   }^2W_1,\partial _{\eta }^2W_2,\partial _{\eta }^3W_2,\partial
   _rW_1,\partial _r\partial _{\eta }W_2,\partial _r^2W_1,\partial
   _r^2\partial _{\eta }W_2\right\} \,,
\ee
with appropriate non zero coefficients depending on $r$ and $\eta$, and on the parameters $h$, $\D$, $l$, $\D_{12}$ and $\D_{34}$.
As a first step, it is natural to try to simplify the system. To do so our criteria was to find new differential equations which involve less terms with high order derivatives. 
For example, if we ask for which coefficients $a(r,\eta)$ the  combination $ \texttt{PDE}_1 + a(r,\eta) \texttt{PDE}_2$ can be written just in terms of
\be
\left\{W_1,\partial _{\eta }W_1,\partial _{\eta }W_2,\partial _{\eta
   }^2W_1,\partial _rW_1,\partial _r^2W_1\right\}
\ee
we obtain that $a(r,\eta)=  -\eta$. Similarly, if we want to use the set
\be
\left\{W_1,\partial _{\eta }W_2,\partial _{\eta }^2W_2,\partial _{\eta
   }^3W_2,\partial _r\partial _{\eta }W_2,\partial _r^2\partial _{\eta
   }W_2\right\}
\ee
we find the condition $ a(r,\eta)=  \left(r^2+1\right)/(2r)$.
Therefore we obtain a simpler system of equation by choosing
\be
\texttt{PDE}_1-\eta \, \texttt{PDE}_2=0  \,, \qquad \qquad
2r \, \texttt{PDE}_1-(r^2+1) \, \texttt{PDE}_2=0
\,.
\ee
Applying this simplification we obtain two coupled algebraic relations $\texttt{AR}_s$ for the coefficients $w_{s}$. The first of which is actually the one defined in \eqref{recrel1Vector} with $s=1$. The second one can instead be schematically written as
 \be \label{recrel1VectorWRONG}
\texttt{AR}_2(m,j) \equiv \sum_{s=1}^2 \sum_{(\hat{m},\hat{j}) \in \tilde\Scal_{s}} \tilde c_{s}(\hat{m},\hat{j}) \; w_{s}(m+\hat{m},j+\hat{j})=0
 \ ,
\ee
where the coefficients $\tilde c_{s}$ are in general different from the ones of \eqref{recrel1Vector} and $\tilde \Scal_{s}$ is represented in figure \ref{fig:RecCoefficient1Vector2WRONG}.
Equation \eqref{recrel1VectorWRONG} can still be used
to find the coefficients of the expansions of $W_1$ and $W_2$ even if there are two points with maximal $\hat m$. In fact, one can for example choose to express $w_s(m,j)$ with maximal $m$ and lower $j$ in terms of the other  coefficients. But this is not enough, one also needs to assume that the expansion of $W_s$ is of the form of figure \ref{fig:Expansion_Plot}. Therefore, at each level $m$ one should start by expressing $w(m,j)$ starting from $j=l+m$ and decreasing $j$ until $\max(0,l-m)$, and imposing that the coefficients with $j \notin [\max(0,l-m),l+m]$ vanish. However, one can convince oneself that the expansion of figure \ref{fig:Expansion_Plot} implies that  figure \ref{fig:RecCoefficient1Vector2WRONG} is not minimal, and that there should exist a minimal  relation which has just a single dot with maximal $\hat m$. A relation with this feature  can actually be obtained as follows. Let us  first define a new set of algebraic relations indexed by an integer $n$,
\be
\texttt{ar}^{(n)}(m,j) =\sum_{k=0}^n \texttt{AR}_2(m,j-1-2 k) \,.
\ee
When $n=0$, the \lq\lq{}dot diagram\rq\rq{} of $\texttt{ar}^{(0)}(m,j)$, would simply be a shifted version of figure \eqref{fig:RecCoefficient1Vector2WRONG}.
\begin{figure}
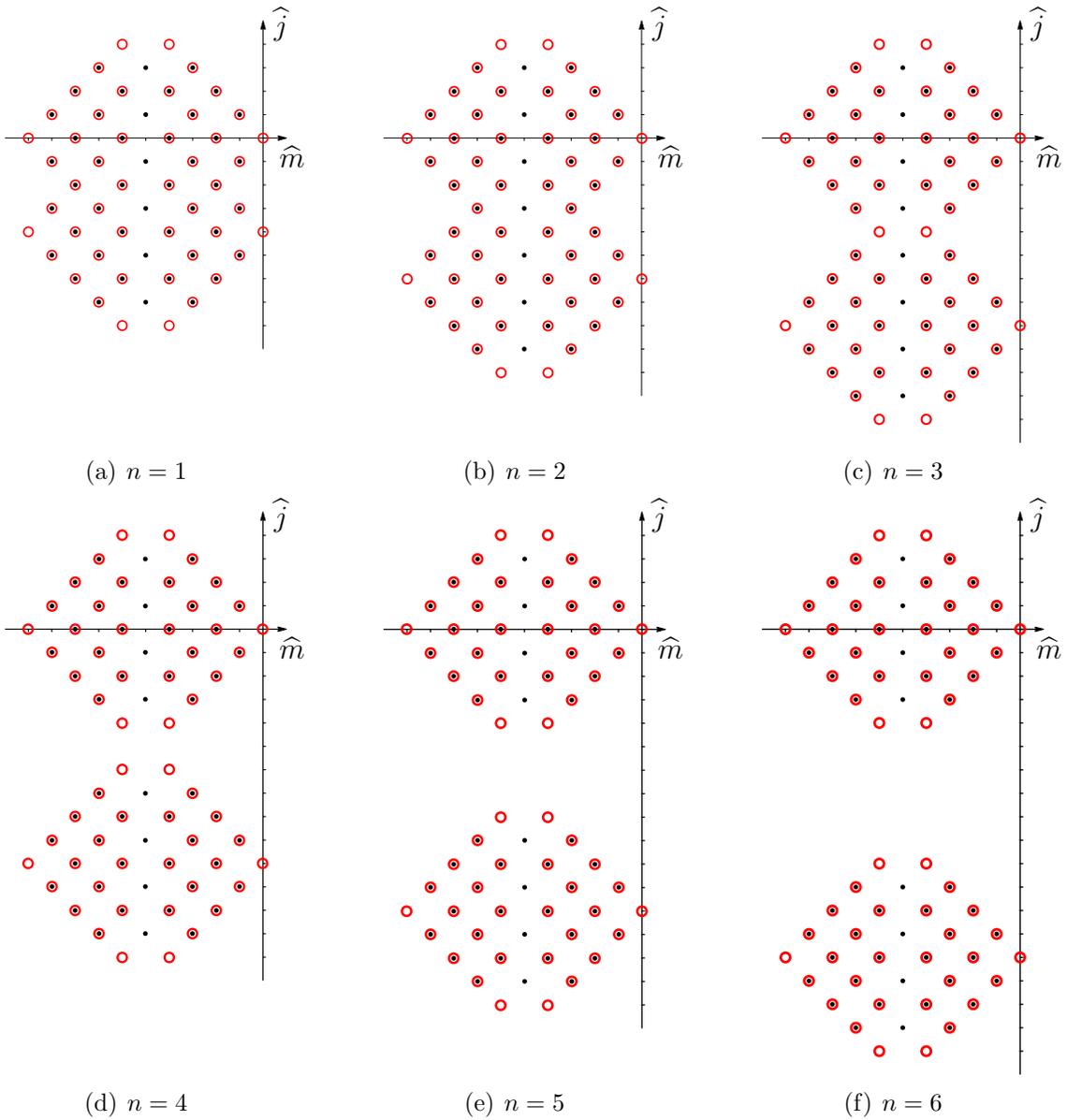

\centering
\subfigure[\label{fig:disentangling1} $n= 1$]{\graphicspath{{Fig/}}
\def\svgwidth{4 cm} 
 \input{disentangling1.pdf_tex}}
~~~~~~
\subfigure[\label{fig:disentangling2} $n= 2$]
{\graphicspath{{Fig/}}
\def\svgwidth{4 cm} 
 \input{disentangling2.pdf_tex}}
~~~~~~
\subfigure[\label{fig:disentangling3} $n= 3$]
{\graphicspath{{Fig/}}
\def\svgwidth{4 cm} 
 \input{disentangling3.pdf_tex}}
~~~~~~
\subfigure[\label{fig:disentangling4} $n= 4$]
{\graphicspath{{Fig/}}
\def\svgwidth{4 cm} 
 \input{disentangling4.pdf_tex}}
~~~~~~
\subfigure[\label{fig:disentangling5} $n= 5$]
{\graphicspath{{Fig/}}
\def\svgwidth{4 cm} 
 \input{disentangling5.pdf_tex}}
~~~~~~
\subfigure[\label{fig:disentangling6} $n= 6$]
{\graphicspath{{Fig/}}
\def\svgwidth{4 cm} 
 \input{disentangling6.pdf_tex}}
\caption{\label{fig:disentangling}The algebraic relation $\texttt{ar}^{(n)}(m,j)$. Increasing the value of $n$ it is possible to disentangle two relations.
}
\end{figure}
 However, increasing the value of $n$, we found (as a result of a computation) that in the \lq\lq{}dot diagram\rq\rq{} of $\texttt{ar}^{(n)}(m,j)$ two distinct shapes of points appear and disentangle one from the other, as pictured in figure \ref{fig:disentangling}. In figure \ref{fig:disentangling} we considered values of $n$ up to  $n=6$, but  increasing $n$ further it is possible to send the lower shape of points at arbitrary large and negative values of $\hat j$. 
When we use $\texttt{ar}^{(n)}(m,j)$ with large enough $n$ as a recurrence relation, we can simply forget about the lower shape of points since they are just combinations  of coefficients $w_s(m,j)$ with $j<0$, which are zero. Since the coefficients of the higher shape do not change by increasing $n$, we can just consider the relation  $\texttt{ar}^{(n)}(m,j)$ for any $n$ with $n\geq 4$ (when the two shapes are really disentangled) and set the lower shape to zero. In this way we obtain the algebraic relation \eqref{recrel1Vector}, which only has one dot with maximal $\hat m$.

When we consider two external vectors more complicated structures can arise but it is enough to use the ideas of this section to simplify the recurrence relations.

\section{Details on the recursion relation from the analytic structure in $\D$}
\label{appendix:recdeltaFULL}
This appendix is devoted to the computation of all the elements which enter formulae \eqref{eq:rechsymtraceless} and \eqref{recrelh00}.
In the first section we compute the coefficients $R_A$ and in the second section we find the large $\D$ behavior of the blocks. Finally, in the last section we perform the leading OPE computation for the blocks in the natural normalization provided by the method of \cite{ arXiv:1509.00428}.

\subsection{The residues $R_A$}
\label{appendix:recdelta}
In this appendix we want to give more details on the computation of all the coefficients $R_A$ in formulas 
\eqref{eq:rechsymtraceless} and \eqref{recrelh00}.
We start by a complete derivation of $(R_{\I_2 , 1})_{p q}$ and $(R_{\II_2 , 1})_{p q}$. Then we explain how to obtain $(R_A)_{pp\rq{} q q\rq{}}$ of \eqref{eq:rechsymtraceless} from the results of \cite{ arXiv:1509.00428}. Finally we write the coefficients $R_A$ of \eqref{recrelh00}, which we obtained by matching the expansion in $r$.

In the following we obtain from firsts principles the coefficients $(R_{\I_2 , 1})_{p q}$ and $(R_{\II_2 , 1})_{p q}$ of \eqref{RI21} and \eqref{RII21}. 
Notice that these coefficients arise from new poles which were predicted in \cite{ arXiv:1509.00428}, but still not observed.
These arise from (null) primary descendant states which are in a different representation with respect to the primaries.
In particular, the primary descendant of type $\I_2 , 1$ is in the representation $(\D+1,l,1)$, and it is a descendant at the level one of a primary state in the representation $(\D,l)$. Conversely, the descendant of type $\II_2 , 1$ lives in $(\D+1,l)$, and is a descendant at the level one of  a primary in $(\D,l,1)$. Schematically
\be
\begin{array}{ccc}
(\, \D,l \,) &\overset{\I_2 , 1}{\longrightarrow}&(\D+1,l,1) \,,\\
(\D,l,1)& \overset{\II_2 , 1}{\longrightarrow}&(\, \D+1,l \,) \,.
\end{array}
\ee
Using the language of \cite{ arXiv:1509.00428}, we generate the  descendant states acting with an operator $\Dcal_A$ on primary states. These descendant states become primary (and therefore null) when the conformal dimension $\D$ of the primaries takes specific values,
\be
\begin{array}{llll}
|\I_2 , 1 \rangle &\equiv \Dcal_{\I_2 , 1}& |(\D , l);z \rangle 
\ ,
&
\mbox{becomes primary when } \D=\D^\star_{\I_2 , 1} \equiv 1 \ ,
\\
|\II_2 , 1 \rangle &\equiv \Dcal_{\II_2 , 1} & |(\D , l ,1);z,w \rangle \ ,
\qquad
&
\mbox{becomes primary when } \D=\D^\star_{\II_2 , 1} \equiv 2h-2 \ .
\end{array}
\ee
The primary states $|(\D , l);z \rangle$ and $|(\D , l ,1);z,w \rangle $ are contracted with polarization vectors as follows
\begin{align}
|(\D , l);z \rangle &\equiv  |(\D , l); \ytableausetup{centertableaux,boxsize=1.2 em}
{ \scriptsize
\begin{ytableau}
\mu_1&\mu_2&\, _{\cdots}&\mu_l \\
\end{ytableau}
}\,
\rangle
z^{\m_1}z^{\m_2} \cdots z^{\m_l} 
\nonumber\\
| (\D , l ,1);z,w \rangle  &\equiv  
\Big|
(\D , l,1); \ytableausetup{centertableaux,boxsize=1.2 em}
{ \scriptsize
\begin{ytableau}
\mu_1&\, _{\cdots}&\mu_l \\
\nu \\
\end{ytableau}
}\,
\Big\rangle
z^{\m_1}\cdots z^{\m_l}w^{\n}
\end{align}
and the new operators $\Dcal_A$ are 
\begin{align}
\begin{split}
\Dcal_{\I_2,1} &=\frac{-2 i}{ (h+l-2) (2 h+l-4)} \; \left(z_{[\n} w_{\m]})(P^{[\n} D_{z,w}^{\m]}\right) , 
\\
\Dcal_{\II_2,1} &=\frac{-i}{(h-2) (h-1)} \;(P \cdot D_{w,z}) \,,
\end{split}
\label{DAops}
\end{align}
where the overall normalizations are chosen for convenience, and $D_{z,w}$ and $D_{w,z}$ are the 
differential operators which generate projectors to traceless mixed-symmetry tensors \cite{projectors}. They are defined as
\be
\label{NewTodorov}
D^{\m}_{z,w}\equiv d_{00} \partial^{\m}_{z}+ d_{-1 1} \partial^{\m}_{w}+z^{\m} d_{-2 0} +w^{\m} d_{-1 -1} \,,
\ee
where $d_{m n}$ are differential operators with weight $m$ in the variable $z$ and $n$ in the variable $w$,
\begin{align} 
\begin{split}
d_{00}& \equiv  (1-h) \big[ (2 h-3)+3  (z \cdot \partial_{z})+  (w\cdot \partial_{w})\big]-(z\cdot \partial_{z})(w\cdot \partial_{w})-z^{\a}(z\cdot \partial_{z}) \partial_{z\, \a}\,,
\\
 d_{-2 0}& \equiv \frac{1}{2}\big[2(h-1)+(w\cdot \partial_{w}) +(z\cdot \partial_{z})\big](\partial_{z}\cdot \partial_{z}) \,,
 \\
d_{-1 1} &\equiv-2 (h-1) (w\cdot \partial_{z})-(w\cdot \partial_{z})(w\cdot \partial_{w})-(z\cdot \partial_{z})(w\cdot \partial_{z})\,,
\\
d_{-1 -1}& \equiv \big[(h-1) +(z\cdot \partial_{z}) \big] (\partial_{z}\cdot \partial_{w})
+\frac{1}{2} \big[ (w\cdot \partial_{z})(\partial_{w}\cdot \partial_{w})- (z\cdot \partial_{w}) (\partial_{z}\cdot \partial_{z})\big] \,,
\end{split}
\end{align}
and similarly for $D^{\m}_{w,z}$ with $z \leftrightarrow w$.

Using the operators $\mathcal{D}_A$ in (\ref{DAops}) we can proceed to the computation of the coefficients $R_A$ as explained in \cite{ arXiv:1509.00428}. The following computations are done in a new normalization, with respect to the main text, which is conveniently chosen to apply the techniques introduced in \cite{ arXiv:1509.00428}. In the end of the appendix we explain how to return to the conventions used in the main text. 

First we compute $Q_A$, the residue at the pole $\D^\star_A$ of the inverse of  norms 
$N_{\I_2 , 1}\equiv \langle \I_2 , 1|\I_2 , 1 \rangle$ and $N_{\I_2 , 1}\equiv \langle \II_2 , 1 |\II_2 , 1 \rangle $,
\be
\frac{1}{N_{A} }\sim \frac{Q_{A}}{\D- \D^\star_A} \ .
\ee
This follows from the commutation relations of the conformal algebra, and the result is 
\be
\label{QI21QII21}
 Q_{\I_2 , 1}=\frac{1}{l (l+1)} \,,\qquad\qquad 
 Q_{\II_2 , 1}=\frac{h-1}{2 (2 h+l-3) (2 h+l-2)} \,.
\ee
Next we need to obtain the relation between the OPE coefficients of the primary descendant and the OPE coefficients of the original primary.
It is convenient to introduce the leading OPE of the three point function of a vector operator $\Ocal_1$, a scalar operator $\Ocal_2$ and an operator $\Ocal$ in the representations $(\D,l)$ and $(\D,l,1)$,
\be
\Ocal(x,z) \Ocal_1(0,z_1)=\frac{\Ocal_2(0)}{(x^2)^{\a_{\D_{12}}}}  \times  \left\{
\begin{array}{l l}
 c^{(0)}_{12\Ocal} t_l^{(0)}(x,z,w,z_1)
& \mbox{  if } \Ocal \in (\D,l,1) \\
c^{(1)}_{12\Ocal} t_l^{(1)}(x,z,z_1)+ c^{(2)}_{12\Ocal} t_l^{(2)}(x,z,z_1) 
& \mbox{  if } \Ocal \in (\D,l)
\\
\end{array}
\right.
\ee
with $\a_{x}\equiv\frac{\D+x+l+1}{2}$ and 
\begin{align}
\label{tpl}
\begin{split}
t_l^{(0)}(x,z,w,z_1)&=\displaystyle \frac{1}{2}|x| (x \cdot z)^{l-1} \big[ (w \cdot z_1) (z \cdot x)-(z \cdot z_1) (w \cdot x)\big]\ , 
\\
 t_l^{(1)}(\, x,z,z_1 \,)&=\displaystyle (x \cdot z)^l (x \cdot z_1) \ , \\ 
t_l^{(2)}(\, x,z,z_1 \,)&=\displaystyle (x \cdot z)^{l-1} x^2 (z \cdot z_1) \ .
\end{split}
\end{align}
Therefore we can define the vectors $ M_{\I_2,1}$ and $ M_{\II_2,1}$ from the following computation \cite{ arXiv:1509.00428}
\begin{align}
\begin{split}
\Dcal_{\I_2,1}  \;  \dfrac{t_l^{(p)}(x,z,z_1)}{(x^2)^{\a_\d}} 
&=
\Big( M_{\I_2,1}(\d)\Big)_p  
\dfrac{t_l^{(0)}(x,z,w,z_1)}{(x^2)^{\a_\d+\frac{1}{2}}} \ , \quad 
\qquad (p=1,2)
\\
\Dcal_{\II_2,1}  \;  \dfrac{t_l^{(0)}(x,z,w,z_1)}{(x^2)^{\a_\d}} 
&=
\displaystyle \sum_{p=1}^2 \Big( M_{\II_2,1}(\d)\Big)_p 
 \dfrac{t_l^{(p)}(x,z,z_1)}{(x^2)^{\a_\d+\frac{1}{2}}} \ .
 \end{split}
\end{align}
The result is
\be
\label{MI21MII21}
 M_{\I_2,1}(\d)= \Big( l, \; 2 \alpha_\d -2 \Big) \,, \qquad\qquad 
 M_{\II_2,1}(\d)=\Big( 2 \alpha_\d -1, \; 2 h-2 \alpha_\d +l-1  \Big) \,.
\ee
The residues are then given by
\begin{align}
\begin{split}
\big(\hat{R}_{\I_2,1}\big)_{pq} &=
  \;  Q_{\I_2,1}\;  
\Big(    M_{\I_2,1}(\D_{12})\Big)_p 
\Big(     M_{\I_2,1}(\D_{34})\Big)_q \Big|_{\D=1} \ ,
\\
\big(\hat{R}_{\II_2,1}\big)_{pq} &=
   Q_{\II_2,1}  \;
\Big(  M_{\II_2,1}(\D_{12})  \Big)_p 
\Big(   M_{\II_2,1}(\D_{34})  \Big)_q \Big|_{\D=2h-2} \ .
\end{split}
\end{align}
However, to obtain \eqref{RI21} and \eqref{RII21} we need to return to the normalization of the conformal blocks used in the main text. 

To relate the normalization of the conformal blocks $g_\l$ of the main text to the normalization of the conformal blocks $\hat g_\l$ in this appendix, it is convenient to compare their leading OPE behavior. To find the leading OPE for $\hat g_\l$, we use the techniques proposed in \cite{ arXiv:1509.00428} which are extended in appendix \ref{Conformal_Block_at_large_Delta}. Meanwhile, the leading OPE of $g_\l$ can be trivially found taking the $O(r^0)$ coefficient of the $r$ expansion according to \eqref{IC2vectorl} and  \eqref{IC2vectorl1}. When  $\l=(\D, l)$ we obtain
\begin{align}
\label{normalizationg(D, l)}
 \hat g^{(p,q)}_{(\D, l),s}(r,\eta)&=\mathfrak a_l \sum_{p\rq{},q\rq{}=1}^2 (\mathfrak m^{(L)}_l)_{p,p\rq{}} (\mathfrak m_l^{(R)})_{q,q\rq{}} g^{(p\rq{},q\rq{})}_{(\D, l),s}(r,\eta) \ .
\end{align}
where
\be
\mathfrak a_l\equiv  \frac{ (2 h-1)_{l-1}}{(-2)^{l-1} (h)_{l-1}} \ ,
\qquad
\mathfrak m^{(L)}_l 
 \equiv
\left(
\begin{array}{cc}
- 1 & 0 \\
- 1 & \frac{1}{l} \\
\end{array}
\right) \ ,
\qquad
\mathfrak m^{(R)}_l \equiv
\left(
\begin{array}{cc}
 1 & 0 \\
 1 & \frac{1}{l} \\
\end{array}
\right) \ .
\ee
Similarly, when $\l=(\D, l,1)$, we have
\be
\label{normalizationg(D, l,1)}
\hat g_{(\D, l,1),s}(r,\eta)=\mathfrak b_l \; g_{(\D, l,1),s}(r,\eta) \ ,
\qquad
\mathfrak b_l \equiv  \frac{\mathfrak a_l}{-2 l (2 h+l-3)} \ . 
\ee
Equations \eqref{normalizationg(D, l)} and \eqref{normalizationg(D, l,1)} give rise to the following definition of the coefficients $(R_A)_{pq}$, 
\begin{align}
\begin{split}
\big(R_{\I_2,1}\big)_{pq} &=
\mathfrak a_l^{-1} \;  Q_{\I_2,1}\; \mathfrak b_l \;
\Big( \big(\mathfrak m^{(L)}_l \big) ^{-1} \;  M_{\I_2,1}(\D_{12})\Big)_p 
\Big(  \big(\mathfrak m^{(R)}_l \big) ^{-1} \;  M_{\I_2,1}(\D_{34})\Big)_q 
 \Big|_{\D=1}
\ ,
\\
\big(R_{\II_2,1}\big)_{pq} &=
\mathfrak b_l^{-1} \;  Q_{\II_2,1} \; \mathfrak a_l \;
\Big(  M_{\II_2,1}(\D_{12}) \; \mathfrak m^{(L)}_l
\Big)_p 
\Big(   M_{\II_2,1}(\D_{34}) \; \mathfrak m^{(R)}_l
\Big)_q 
\Big|_{\D=2h-2} \ ,
\end{split}
\end{align}
which finally gives \eqref{RI21} and \eqref{RII21}, once we replace the values of $ Q_A$ and $ M_A$ computed in \eqref{QI21QII21} and \eqref{MI21MII21}.

Similarly, we obtain the coefficients  $(R_A)_{pp\rq{} q q\rq{}}$ of \eqref{eq:rechsymtraceless}, 
\be
(R_A)_{pp\rq{} q q\rq{}} = \mathfrak a^{-1}_{l} \;  Q_{A} \; \mathfrak a_{l_A} \;  \Big( \big(\mathfrak m^{(L)}_{l} \big)^{-1} \;  M_{A}^{(L)} \; \mathfrak m^{(L)}_{l_A} \Big)_{p p\rq{}}
 \;  \Big(\big(\mathfrak m^{(R)}_{l} \big)^{-1}\;  M_{A}^{(R)} \; \mathfrak m^{(R)}_{l_A} \Big)_{q q\rq{}}  \Big|_{\D=\D^\star_A}\,,
\ee
where $ Q_{A}$ and $ M_{A}$ are presented in formulae (34-36) and (189) of  \cite{ arXiv:1509.00428}.

The coefficients $R_A$ of \eqref{recrelh00}
 could in principle be computed using the same technology that we used in this appendix, but we decided to find them matching terms in the expansion in $r$ of the conformal blocks. This can be easily done since these coefficients do not present any matrix structure. Moreover, they are not conceptually new since they arise from null states created by the operators $\Dcal_{\I,n}, \Dcal_{\II,n}$ and $\Dcal_{\III,n}$ already defined in \cite{ arXiv:1509.00428}. The result is
\begin{align}
R_{\I_1,n}&=-\frac{l  \left(\frac{-n+\Delta_{12}+1}{2}\right)_n \left(\frac{-n+\Delta_{34}+1}{2} \right)_n (2 h+l-3)_n}{(n-1)! n! (l+n) (h+l-1)_n} \,,
\nonumber\\
R_{\II_1,n}&=
-\frac{(-1)^n (-l-1)_n \left(\frac{-n+\text{$\Delta $12}+1}{2} \right)_n \left(\frac{-n+\text{$\Delta $34}+1}{2} \right)_n (2 h+l-n+1-2)_n}{n! (n-1)! (h+l-n+1-1)_n (2 h+l-n+1-3)_n}\,,
\label{RA(D,l,1)}
\\
R_{\III,n}&=\frac{ \left(\frac{h-n+1}{2}\right)_n \left(\frac{h-n-2}{2}\right)_n \left(\frac{\Delta _{12}+h+l-n}{2}\right)_n \left(\frac{\Delta_{12}-h-l-n+2}{2}\right)_n \left(\frac{\Delta_{34}+h+l-n}{2}\right)_n \left(\frac{\Delta _{34}-h-l-n+2}{2}\right)_n}{(-1)^{n+1} 4^{-n} n! (n-1)! (h+l-n)_{2 n} (h+l-n-1)_{2 n}}
\,.
\nonumber
\end{align}

\subsection{Conformal Block at Large $\D$}
\label{ConformalBlockAtLargeDelta}

The goal of this section is to find
$h^{(p,q)}_{(\infty,l),s}(r,\eta)$ of \eqref{eq:rechsymtraceless} and
 $h_{(\infty,l,1),s}(r,\eta)$ of  \eqref{recrelh00}.
They can be obtained solving the leading term in $\D$ of the Casimir equation 
\be
\mathcal{C} \; \sum_{p,q} c^{(p)}_{12 \Ocal} c^{(q)}_{34 \Ocal}G_{\l}^{(p,q)}(\{P_i,Z_i\})= c_{\l} \sum_{p,q} c^{(p)}_{12 \Ocal} c^{(q)}_{34 \Ocal}G_{\l}^{(p,q)}(\{P_i,Z_i\}) \label{casimir eq spin} \ .
\ee
First we replace \eqref{GTog_Generic} and $g_{\l,s}(r, \eta)=(4r)^{\D} h_{\l,s}(r, \eta)$ in  (\ref{casimir eq spin}), and then we keep the term linear in $\D$. We obtain five coupled first order differential equations in the variable $r$ for the functions $h_{\infty,\l,s}(r, \eta)\equiv \lim_{\Delta \to \infty} h_{\l,s}(r, \eta)$.
 We then make the ansatz for the function $h_{\infty,\l,s}$,
\be 
\label{hss}
h_{\infty,\l,s}(r, \eta)=\Acal^{\D_{12},\D_{34}}(r,\eta) \sum_{t=1}^5 F_{s}^{\,t}(r,\eta)
h_{\l,t}(0,\eta)  \ ,
\ee
where
\be
\Acal^{\D_{12},\D_{34}}(r,\eta)= \frac{ \left(1-r^2\right)^{-h-1}}{ \left(1+r^2-2 r \eta \right)^{\frac{-\D_{12}+\D_{34}+1}{2}} \left(1+r^2+2 r \eta \right)^{\frac{\D_{12}-\D_{34}+3}{2}} }  \ , 
\label{AnsatzLargeDelta}
\ee
and $F_{s}^{\,t}(0,\eta)=\d_{s,t}$ so that $h_{\infty,\l,s}(0,\eta)=h_{\l,s}(0,\eta) $. Using the ansatz (\ref{hss}), the five coupled  differential equations  can be schematically written as
\begin{align}
 \label{2CoupledPDEs}
\partial_r F_s ^{\,t}(r,\eta)&= \sum_{s\rq{}=1}^{5} \mathcal{M}_{s\, s\rq{}}(r,\eta) F_{s\rq{}}^{\;t}(r,\eta) \ ,
\end{align}
where $\mathcal{M}_{s\, s\rq{}}$ is a known $5\times 5$ matrix with entries dependent on $r$ and $\eta$.
The label $t=1,\dots 5$ of $F_s ^{\,t}$ parametrize the five  independent solutions of \eqref{2CoupledPDEs}.
Even if solving a generic system of five coupled first order differential equations is usually very difficult, in this case we can easily obtain the full solution. This is possible since the choice of the ansatz \eqref{AnsatzLargeDelta} turns the solution $F_s^{\,t}$ into  polynomials of $r$ and $\eta$ that can therefore be obtained by expanding \eqref{2CoupledPDEs} in series of $r$, and then solving it term by term. The choice of \eqref{AnsatzLargeDelta}, which makes this possible, is an educated guess inspired by the  large $\D$ behavior of the scalar conformal block.
The result is
\be
F(r,\eta)=\left(
\begin{array}{ccccc}
 \left(r^2-1\right)^2 \left(2 r \eta +A_3\right) & 0 & 0 & 0 & 0 \\
 -2 r^2 A_1^2 & -A_1 A_2 A_3 & -2 r A_1 A_3 & 2 r^2 \eta  A_1 A_2 & 4 r^3 \eta  A_1 \\
 -2 r \left(2 \eta  r^3-r^2-1\right) A_1 & -2 r^2 \eta  A_1 A_3 & A_1 A_3^2 & 4 r^4 \eta ^2 A_1 & -2 r^2 \eta  A_1 A_3 \\
 2 r \left(3 r^2-2 \eta  r-1\right) A_1 & 2 r A_1 A_2 & 4 r^2 A_1 & A_1 A_2^2 & 2 r A_1 A_2 \\
 -2 r^2 A_1^2 & 4 r^3 \eta  A_1 & -2 r A_1 A_3 & 2 r^2 \eta  A_1 A_2 & -A_1 A_2 A_3 \\
\end{array}
\right),
\ee
where 
\be
A_1\equiv r^2-2 \eta  r+1\ , \qquad A_2\equiv r^2-2 \eta  r-1 \ ,\qquad A_3\equiv r^2+1 \ .
\ee
Notice that the matrix $F_{s}^{\,t}$ is independent of the $SO(d)$ representation  of the exchanged operator. This is not surprising since this information  only appears in the eigenvalue $c_\l$ in  terms of order $O(\D^0)$, which do not contribute at leading order at large $\Delta$. 

We can finally obtain
$h^{(p,q)}_{(\infty,l),s}(r,\eta)$ of \eqref{eq:rechsymtraceless}, and
 $h_{(\infty,l,1),s}(r,\eta)$ of  \eqref{recrelh00},
simply by taking the appropriate initial conditions for the functions $W_s$, which correspond to the blocks $G^{(p,q)}_{\D, l }$ and $G_{\D, l , 1}$.

\subsection{The OPE limit of Conformal Blocks}
\label{Conformal_Block_at_large_Delta}
In this appendix  we explain how to obtain the leading behavior of the conformal blocks for $r \rightarrow 0$,  for the exchange of a generic representation $\l$ in the normalization   used in  \cite{ arXiv:1509.00428}, which here we denote $\hat  G_{\l}^{(p,q)}$. To do so, we need to introduce some generic definitions.
The leading order (in $x_{12}$) OPE between two operators  $\Ocal_1$ and $\Ocal_2$, with conformal dimensions $\D_1$ and $\D_2$, and spins $l_1$ and $l_2$, is 
\be \label{OPEwithSPIN1}
\Ocal_1(x_1,z_1)\Ocal_2(x_2,z_2)\approx  \frac{\Ocal(x_{2},{\bf Y})}{(x_{12}^2)^{\frac{\D_1+\D_2-\D}{2}}}\sum_{q} c_{12\Ocal}^{(q)}  t^{(q)}
\big(- \hat x_{12},{\bf Y},I(x_{12})\cdot z_1,z_2 \big) \,,
\ee
where $\Ocal$ is a primary labeled by the conformal dimension $\D$ and by a Young tableau ${\bf Y}$ of $SO(d)$, which has at most $[h]$ lines
 and with  $\ell_i$ boxes in the $i$-th line filled with the indices $\m^{(i)}_1, \dots \m^{(i)}_{\ell_i}$. The notation  $\hat x^\m $ stands for $x^\m/|x|$ and the index $q$ of $t^{(q)}(x,{\bf Y}, z_1, z_2)$ labels the  possible  tensor structures generated by $z_i^{\m}$ and $x^{\m}$ such that 
\be
t^{(q)}(\a x,{\bf Y},\a_1 z_1,\a_2 z_2)=\a^{\ell_1 +l_1+l_2 } \a_1^{l_1} \a_2^{l_2} t^{(q)}(x,{\bf Y}, z_1, z_2) \ .
\ee 
As an example the structures $t^{(q)}$ defined in \eqref{tpl} are of this form, where the indices of ${\bf Y}$ are contracted with polarization vectors $z$ and $w$.  When the set of indices ${\bf Y}$ appears twice it means that the respective indices are contracted
\be
\mathcal{T}({\bf Y}) \Tcal\rq{}({\bf Y}) \equiv \Tcal^{\m^{(1)}_1  \dots \, \m^{(1)}_{\ell_1}  \dots \, \m^{([h])}_1  \dots\, \m^{([h])}_{\ell_{[h]}} }
 \Tcal\rq{}_{\m^{(1)}_1  \dots\, \m^{(1)}_{\ell_1} \dots \,   \m^{([h])}_1  \dots \,\m^{([h])}_{\ell_{[h]} } }\ .
\ee
We shall denote by $\pi({\bf Y},{\bf Y\rq{}})$ the projector  on the Young tableau labeled by ${\bf Y}$.  In this notation a two point function of generic operators can be naturally normalized as
\be
\label{generic2pt}
\langle \Ocal(x_1,{\bf Y}) \Ocal(x_2,{\bf Y\rq{} }) \rangle = \frac{ \pi \big(I(x_{12}) \cdot {\bf Y}, {\bf Y\rq{}}\big) }{(x_{12}^2)^{\D}} \ ,
\ee
where $I(x_{12}) \cdot {\bf Y}$ means that we contract one tensor $I_{\m \n}(x_{12})$ with each index of the young tableau ${\bf Y}$ and where $I_{\m \n}(x)\equiv\d_{\m \n}-2 \hat x_{\m} \hat x_{\n}$.

We consider the four point function of operators $\Ocal_i$ with conformal dimensions  $\D_i$ and spin $l_i$, and repeat the reasoning of \cite{ arXiv:1509.00428}. Taking the leading OPE (\ref{OPEwithSPIN1}) in the channels $1-2$ and $3-4$, and  writing the remaining two point function as \eqref{generic2pt}, we obtain
\begin{align} \label{formula:CBsmallr1}
\hat  G_{\D, l}^{(p,q)} 
&\approx  \pi \big( I(x_{24}) \cdot {\bf Y\rq{}}, {\bf Y } \big) \;  \frac{t^{(p)} \big(\hat x_{12},{\bf Y\rq{}} ,I(x_{12}) \! \cdot \! z_1,z_2\big)t^{(q)}\big(\hat x_{34},{\bf Y} ,I(x_{34}) \! \cdot \! z_3,z_4\big)}{ (x^2_{12})^{\frac{\D_1+\D_2-\D}{2}}(x^2_{34})^{\frac{\D_3+\D_4-\D}{2}}(x^2_{24})^\D }\,.
\end{align}
At  the leading order in the limit $x_{12},x_{34}\rightarrow 0$ we have 
\be
x_{13}\approx x_{23} \approx x_{2 4} \approx x_{14}\ , \qquad (4 r)^2 \approx \frac{x_{12}^2 x_{34}^2}{(x_{24}^2)^2} \ .
\ee
In this limit, we obtain the following formula 
\begin{align}
\hat  G_{\l}^{(p,q)} 
\approx (4 r)^{\D}  \,\frac{ t^{(p)} \big( \hat x_{12},I(x_{24})\cdot {\bf Y},I(x_{12}) \cdot  z_1,z_2 \big)t^{(q)}\big(\hat x_{34},{\bf Y'},I(x_{34}) \cdot z_3,z_4\big)}{ (x^2_{12})^{\frac{\D_1+\D_2}{2}} \qquad (x^2_{34})^{\frac{\D_3+\D_4}{2}}  } \,
\pi({\bf Y},{\bf Y\rq{}})
 \ . \label{smallrCBspin}
\end{align}
On the other hand the OPE limit of (the hatted version of) \eqref{CB:structuresEmbedding} is
\be 
\label{hatLeadOPECB}
\hat G_{\l}^{(p,q)}=\frac{ (4r)^\D}{(x^2_{12})^{\frac{\D_1+\D_2}{2}}(x^2_{34})^{\frac{\D_3+\D_4}{2}}} \sum_{s} \hat h^{(p,q)}_{\l,s}(r, \eta) Q^{(s)} \ ,
\ee 
 where we used $\hat g^{(p,q)}_{\l,s}=(4 r)^\D \hat h^{(p,q)}_{\l,s}$.
We can finally compare \eqref{hatLeadOPECB} with  (\ref{smallrCBspin})  which gives the main formula to define $ \hat h^{(p,q)}_{\l,s}(0,\eta)$,
\begin{align} 
\begin{split}
\label{FinalLeadOPE}
t^{(p)} \big( \hat x_{12}, I(x_{24})\!\cdot \! {\bf Y} ,I(x_{12}) \! \cdot \! z_1,z_2\big)
   t^{(q)}\big(\hat x_{34},{\bf Y'},I(x_{34})\! \cdot \!z_3,z_4\big) 
   \pi({\bf Y},{\bf Y\rq{}})
\\
\approx  \sum_{s} \hat h^{(p,q)}_{\l,s}(0,\eta) Q^{(s)}
\,.
\end{split}
\end{align}
To actually compare the two sides of equation \eqref{FinalLeadOPE} we need to expand the left hand side in the structures $Q^{(s)}$ using the fact that   ${V}_{i,jk}$, ${H}_{i j}$ and $\eta$, for small $r$, behave as 
\be
\begin{array}{c l  l  c l l}
{V}_{a,b i}&\approx& z_a \cdot \hat x_{a b} \ , 
& {V}_{i,j a}&\approx& z_i \cdot \hat x_{i j} \ ,\\
{V}_{i,1 2}&\approx& - z_i \cdot I(x_{2 4})\cdot \hat x_{12} \ , &
{V}_{a,3 4}&\approx& -z_a \cdot I(x_{2 4}) \cdot \hat x_{34} \ ,\\
{H}_{12}&\approx& z_1 \cdot I(x_{12}) \cdot z_2  \ ,\qquad \qquad 
&{H}_{34}&\approx& z_3\cdot I(x_{34}) \cdot z_4  \ , \\
{H}_{a i}&\approx& z_a \cdot I(x_{2 4}) \cdot z_i  \ ,
& \eta &\approx& - \hat x_{12} \cdot I(x_{24}) \cdot \hat x_{34} \ ,
\end{array} \label{smallr:VH}
\ee
where $a,b\in \{1,2 \}$ and $i,j \in \{3,4\}$.
The only new part of \eqref{FinalLeadOPE} with respect to what was discussed in \cite{ arXiv:1509.00428} is that we have to implement the contraction of the indices ${\bf Y}$. To do so we need to know the projector into the representation labeled by ${\bf Y}$.

In the case of the four point function with two vector operators at the points $x_1$ and $x_3$ and two scalars, we know how to generate the projectors in all the exchanged representations, therefore we can use equation \eqref{FinalLeadOPE} to obtain the leading OPE of the conformal blocks. 
For the case of the symmetric and traceless exchange, formula \eqref{FinalLeadOPE} simply reduces to what was already discussed in \cite{ arXiv:1509.00428}. Therefore, one can easily obtain $\hat g_{(\D,l),s}^{(p,q)}(0,\eta)$ from the definitions of $ t^{(1)}_l$ and $ t^{(2)}_l$ in \eqref{tpl}. 
Instead, the case of the  exchange of $(\D,l,1)$ is new.  
We can implement the contraction of the indices of \eqref{FinalLeadOPE} in two ways, either using the operators $D_{z,w}$, $D_{w,z}$ of \eqref{NewTodorov}, or using the closed form for the projector proposed in \eqref{Vmunu}.
Following the first way we would obtain
\begin{align}
\begin{split}
\Ncal_{l,1} \;  t^{(0)}_l\big( \hat x_{12}, 
   I(x_{24}) \! \cdot \! D_{z,w},I(x_{24})\! \cdot \! D_{\omega,z} ,I(x_{12}) \! \cdot \!   z_1\big) \; t^{(0)}_l\big(\hat x_{34},z,\omega,I(x_{34})\! \cdot \!  z_3\big)  
   \\
 \approx  \sum_{s} \hat h_{(\D,l,1),s}(0,\eta) Q^{(s)} \ ,
\end{split}
\end{align}
where $ t^{(0)}_l$ is defined in \eqref{tpl}. Here we introduced the factor  $\Ncal_{l,1}\equiv \frac{4 l (-1)^{-l-1}}{(l+1)! (h-1)_l (2h-4)_{l+2}}$ in order to properly normalize the  projector into the $(l,1)$ irrep, as shown in \cite{projectors}. Equivalently, we can use the closed form representation  \eqref{Vmunu} of the double vector harmonic 
\be
\big( I_\m^{\; \b}(x_{24}) I_{\b \a}(\hat x_{12}) z_1^{\a} \big)  \big(I_{\n \d}(\hat x_{34}) z_3^{\d}\big) \Vcal^{\m  \n} \big(I(x_{24})\cdot \hat x_{12}, \hat x_{34}\big)
 \approx  \sum_{s} \hat h_{(\D,l,1),s}(0,\eta) Q^{(s)} \,.
\ee
From these formulas one can completely fix $\hat h_{(\D,l,1),s}(0,\eta)$  as
\begin{align}
\begin{split}
&\hat h_{(\D,l,1),2}(0,\eta)=\hat h_{(\D,l,1),5}(0,\eta)=c_l \;  C_{l-2}^{(h+1)}(\eta ) \,, \\
& \hat h_{(\D,l,1),4}(0,\eta)=\frac{2 (1-h)}{l-1}  \,c_l  \; C_{l-3}^{(h+1)}(\eta ) -\frac{(-2 h-l+3)}{l-1} \, \hat h_{(\D,l,1),3}(0,\eta) \,,\\
&\hat h_{(\D,l,1),3}(0,\eta)=\eta \;\hat h_{(\D,l,1),2}(0,\eta) \,,\\
&\hat h_{(\D,l,1),1}(0,\eta)=\eta \;\hat  h_{(\D,l,1),4}(0,\eta)- \hat h_{(\D,l,1),2}(0,\eta)\,,
\end{split}
\end{align}
where $c_l=-\frac{(-2)^{1-l} (l-1)!}{(2 h+l-3) (h+1)_{l-2}}$.

\section{Differential Operators Method}
\label{Spinning_CB}
In this appendix we implement the method of \cite{arXiv:1109.6321} to match some of the results obtained from the expansions in radial coordinates presented in the main text. 
The main idea of \cite{ arXiv:1109.6321} is that 
we can obtain conformal blocks for external operators with spin, by acting with differential operators on the scalar conformal block. Schematically one can write
\be
G_{\D,l}^{(p,q)}(P_i,Z_i)= D_{\textrm{Left}}^{(p)} D_{\textrm{Right}}^{(q)} G_{\D,l}(P_i)\,,
\label{SinningCBgeneric}
\ee
where the operators $ D_{\textrm{Left}}^{(p)}$ and $ D_{\textrm{Right}}^{(q)} $ are explicitly defined in \cite{ arXiv:1109.6321} in terms of multiplication and derivatives  using the embedding vectors $P_i$ and $Z_i$, and by some shifts of the external dimensions.
Using formula \eqref{SinningCBgeneric} one can obtain all the blocks for the exchange of a symmetric and traceless representation. Unfortunately, with this method it is not possible to obtain the blocks for more complicated exchanges. However, one can pursue the philosophy of acting with some differential operators on simple \seed blocks, which exchange a representation $\l$ in order to obtain the most generic blocks labeled by the exchange of the same $\l$.
This idea was fruitfully followed in four dimensions in  \cite{arXiv:1601.05325}.

It is instructive to write explicitly how the differential operators look in radial coordinates. To do so we first  express the scalar block using equation \eqref{GcalTog} and then collect the coefficients multiplying each tensor structure in the  four point function according to \eqref{CB:structuresEmbedding}.
In the case of the conformal block decomposition \eqref{vectorCB} of the four point function with one external vector, we find
\be
\label{hpSpinningCB}
h^{(p)}_{(\D,l),s}(r, \eta)=
\sum_{q=1}^2 (\Mcal)_{pq} \Dcal^{(q)}_s\; h_{(\D,l)}(r,\eta)
 \ ,
\ee
where 
\be
\Mcal=\frac{-1}{2 (\Delta -1)} \left(
\begin{array}{cc}
 1 & 1 \\
1-\Delta - \Delta _{12} \; & \Delta -1- \Delta _{12} \\
\end{array}
\right)
\ee
and the functions $h_{(\D,l)}$ are related to the usual conformal blocks as follows $g_{(\D,l)}= (4r)^\D h_{(\D,l)}$ (and similarly for $h^{(p)}_{(\D,l),s}$). The matrix $\Mcal$ is introduced in order to match the normalization of the conformal blocks that we used in the main text. The operators $\Dcal_s^{(p)}$ are defined as follows
\begin{align}
\begin{split}
\Dcal_1^{(1)} h=
&\left(\frac{2 \left(1+r^2\right) \left(1+\Delta _{12}\right)}{1+r^2+2 r \eta
   }+\frac{1+\Delta +r^2 \left(-1+\Delta -\Delta _{12}\right)+\Delta
   _{12}}{-1+r^2}\right) h^{\left[1\right]}  \\
&-\eta  \partial
   _{\eta }h^{\left[1\right]}+\frac{\left(r+r^3\right) \partial
   _rh^{\left[1\right]}}{-1+r^2} \ ,\\
\Dcal_2^{(1)} h=
&\left(-\frac{2 r \Delta }{-1+r^2}-\frac{4 r \left(1+\Delta
   _{12}\right)}{1+r^2+2 r \eta }\right) h^{\left[1\right]}-\partial _{\eta }h^{\left[1\right]}-\frac{2 r^2
   \partial _rh^{\left[1\right]}}{-1+r^2} \ , \\
\Dcal_1^{(2)} h=&
\frac{\left(1+r^2-2 r \eta \right) \big(-1+\Delta +r^2 \left(1+\Delta
   -\Delta _{12}\right)+\Delta _{12}\big) h^{\left[-1\right]}}{\left(-1+r^2\right) \left(1+r^2+2 r \eta
   \right)}  \\
&+\frac{\eta  \left(1+r^2-2 r \eta \right) \partial _{\eta
   }h^{\left[-1\right]}}{1+r^2+2 r \eta
   }+\frac{\left(r+r^3\right) \left(1+r^2-2 r \eta \right) \partial
   _rh^{\left[-1\right]}}{\left(-1+r^2\right) \left(1+r^2+2 r
   \eta \right)} \ , \\
\Dcal_2^{(2)} h=&
\frac{2 r \big(2 (1+\Delta _{34}-\Delta
   _{12} ) (1-r^2)  -
 \Delta (r^2+1-2 r  \eta) \big) h^{\left[-1\right]}}{\left(-1+r^2\right)
   \left(1+r^2+2 r \eta \right)}  \\
&+\left(-1+\frac{2 \left(1+r^2\right)}{1+r^2+2
   r \eta }\right) \partial _{\eta }h^{\left[-1\right]}-\frac{2
   r^2 \left(1+r^2-2 r \eta \right) \partial _r h^{\left[-1\right]}}{\left(-1+r^2\right) \left(1+r^2+2 r \eta \right)} \ ,
\end{split}
\end{align}
where $h ^{\left[k\right]}$ means that  the variable $\D_{12}$ in the function $h$ is shifted by $k$, according to  $\D_{12} \rightarrow \Delta_{12} + k$. Formula \eqref{hpSpinningCB} is in agreement with the radial expansion of section \ref{subsec:1extvec} and with the results of \cite{ arXiv:1509.00428}.

Similarly, for the case of two external vectors we can write
$
h^{(p,q)}_{(\D,l),s}(r, \eta)= \Dcal^{(p,q)}_s\; h_{(\D,l)}(r,\eta) 
$.
In this case we do not present the explicit action of the operators since it would be too long. However, we define them in a \Mathematica file included with the submission. It is worth mentioning that the result matches the computation of the symmetric and traceless blocks $g^{(p,q)}_{(\D,l),s}(r, \eta)$ obtained both from the $r$ expansion and from the analytic structure in $\D$.

\bibliographystyle{./utphys}
\bibliography{./mybib}

\end{document}